%% file: main_arxiv.tex
\documentclass[journal]{IEEEtran}
\usepackage[utf8]{inputenc}
\usepackage{amsmath}
\usepackage{amsfonts}
\usepackage{amssymb}
\usepackage{array} % Include this in the preamble for better control over column formatting
\usepackage{makecell}
\usepackage{cellspace}
\usepackage{multirow}

\usepackage{graphicx}
\usepackage{hyperref}
\usepackage{cite}
\usepackage{algorithm}
\usepackage{lmodern}
\usepackage{amsmath}
\usepackage{amsfonts}
\usepackage{color}
\usepackage{bm}
\usepackage{graphicx}
\usepackage{algorithm}
\usepackage{multirow}
\usepackage{cite}
\usepackage[table,xcdraw]{xcolor}
\usepackage{comment}

\usepackage{amsmath}
\usepackage{amssymb}
\usepackage{amsthm}
\usepackage{subfigure}
\usepackage{textcomp}
\usepackage{xcolor}
\usepackage{booktabs}
\usepackage{url}

\usepackage{listings}
\usepackage{lipsum}  % For filler text (optional)

\usepackage{algpseudocode}
\usepackage{color}

\usepackage{titletoc}
\usepackage{titlesec}
\hypersetup{
  hidelinks,
  colorlinks=false
}

\usepackage{array}
\newcommand{\PreserveBackslash}[1]{\let\temp=\\#1\let\\=\temp}
\newcolumntype{C}[1]{>{\PreserveBackslash\centering}p{#1}}

\title{The Unseen AI Disruptions for Power Grids: LLM-Induced Transients}

\author{
    \IEEEauthorblockN{Yuzhuo Li, Mariam Mughees, Yize Chen, Yunwei Ryan Li}\\
    \IEEEauthorblockA{Department of Electrical and Computer Engineering\\
    University of Alberta, Edmonton, Canada\\
    \footnotesize{\{yuzhuo, mughees, yize.chen, yunwei.li\}@ualberta.ca}}
}

\begin{document}

\maketitle

\begin{abstract}
Recent breakthroughs of large language models (LLMs) have exhibited superior capability across major industries and stimulated multi-hundred-billion-dollar investment in AI-centric data centers in the next 3-5 years. This, in turn, bring the increasing concerns on sustainability and AI-related energy usage. However, there is a largely overlooked issue as challenging and critical as AI model and infrastructure efficiency: the disruptive dynamic power consumption behaviour. With fast, transient dynamics, AI infrastructure features ultra-low inertia, sharp power surge and dip, and a significant peak-idle power ratio. The power scale covers from several hundred watts to megawatts, even to gigawatts. These never-seen-before characteristics make AI a very unique load and pose threats to the power grid reliability and resilience. To reveal this hidden problem, this paper examines the scale of AI power consumption, analyzes AI transient behaviour in various scenarios, develops high-level mathematical models to depict AI workload behaviour and discusses the multifaceted challenges and opportunities they potentially bring to existing power grids. Observing the rapidly evolving machine learning (ML) and AI technologies, this work emphasizes the critical need for interdisciplinary approaches to ensure reliable and sustainable AI infrastructure development, and provides a starting point for researchers and practitioners to tackle such challenges.
\end{abstract}

%\tableofcontents

\section{Introduction}
\input{intro}

\section{Characteristics of AI Loads}
\input{Characteristics}

\section{Power Consumption Modeling of AI Workloads}
\input{Modeling}

%\section{Challenges for Power Grids}
%There are some overlaps between this section and the future, so for the current version, I decided to blend these into future discussions.
%\input{Challenges}

\section{AI Power Consumption Analysis: Case Studies}
\input{simulation}

\section{Future Research Directions}
\input{Future_and_Opportunities}

\section{Conclusions}
%[Revise the conclusion to incorporate insights from the case study]

%This comprehensive framework addresses the unique challenges posed by AI infrastructure as a load on power grids. By integrating advanced modeling techniques, optimization strategies, and novel technologies, a roadmap for managing the dynamic and intensive power demands of AI systems is provided. The work highlights the need for interdisciplinary collaboration between power systems engineers, computer scientists, and AI researchers to develop sustainable solutions for the growing energy demands of AI. Future work should focus on real-world implementations and further refinement of the proposed methods to ensure the sustainable growth of AI technologies in harmony with our power infrastructure. As AI continues to advance and permeate various aspects of society, the effective management of its energy consumption will be crucial for sustainable development and the reliability of our power systems. The framework and strategies presented in this paper provide a solid foundation for addressing these challenges and pave the way for future innovations in this critical area.

Through comprehensive analysis and investigation, this paper reveals the AI infrastructure as a unique load on power grids. Unlike conventional data center loads, AI, especially LLMs, exhibits extreme dynamics that have never been seen before. Its ultra-low inertia, subsecond peak-idle power fluctuation, unpredictable interruptions during training, and user-behaviour dominant inference power consumption all together pose great challenges to power grids. As AI continues to advance and permeate various aspects of society, the effective management of its power consumption will be crucial for sustainable development and the reliability of our power systems. The analysis, and case study presented in this paper provide a start for addressing these challenges and pave the way for future innovations in this critical area.

\bibliographystyle{IEEEtran}
\bibliography{references}

% Set appendix numbering format to alphabetic
\titleformat{\section}
  {\normalfont\Large\bfseries}
  {Appendix \thesection:}{1em}{}
\renewcommand{\thesection}{\Alph{section}}

\section*{Appendix}
\input{Appendix}

\end{document}

%% file: intro.tex
Recent years have witnessed the explosive growth of AI applications, from edge devices to large-scale data centers. Large language models (LLMs) such as GPT-4, Llama 3, and BERT have largely pushed the boundaries of modern AI~\cite{bommasani2021opportunities, bubeck2023sparks, dubey2024llama}. While such technical advancements are at the cost of significantly increased power and energy consumption. Over the past five years, AI workloads have undergone significant transitions, driven by rapid advancements in AI-related computing. The 2024 electricity report from the International Energy Agency (IEA) estimated that data centers consumed 460 TWh in 2022, accounting for 2\% of global electricity usage-a figure that is still rapidly increasing \cite{iea2024electricity}. Even before the wide adoption of LLMs and Generative AI (GenAI) models~\cite{bommasani2021opportunities, touvron2023llama, achiam2023gpt}, the size of recommendation models at Meta has increased by 20× into the terabyte scale, while inference requests more than doubled from 2019 to 2022, leading to a 2.5× increase in infrastructure capacity \cite{lui2021understanding}. Moreover, energy consumed in both the training and operational stages of GenAI models are surging over the last few years, raising the concerns over AI models' scalability and sustainability~\cite{kaack2022aligning}. For instance, the training of GPT-4 consumed over 50 GWh, approximately 0.02\% of California's annual electricity consumption, representing a 50-fold increase over the energy required to train GPT-3 \cite{forbes2024gpt}. However, focusing solely on these figures might obscure a more critical challenge: the disruptive nature of AI-induced impacts on power grids. Such phenomenon can not only demand far more electricity energy generated and dispatched by power systems, it also significantly affects the reliable and sustainable operation of power grids through the transient load behaviors, the sheer volume of electricity load related to AI computing, and the fast deployments of emerging AI services. If without careful planning and power management, large-scale AI models in both training and deployment stages can cause peak loading and burden the local distribution system. Just for a concrete showcase, as our experiment indicates, LLM training can instantly move from cold start to peak load with over tens of Megawatts of power. While for current existing distribution grid design, such minute-level ramping capability is typically limited to few Megawatts~\cite{kirby2005method}.

\subsection{AI Growth and Associated Impacts on Power Consumption}

The AI industry's energy demand is growing at an alarming rate, with the computational requirements of AI models doubling every 10 months \cite{sevilla2022compute, masanet2024better}. This exponential growth translates into a corresponding increase in energy consumption \cite{crawford2024generative}. When comparing energy consumption across different sectors, it is evident that AI workloads are becoming increasingly significant. %For example, Bitcoin mining alone consumes around 130 TWh per year, while GPT-3.5’s energy consumption is approximately 0.03\% of Bitcoin’s annual usage \cite{vincent2024electricity}. 
In terms of household energy usage, GPT-3.5's annual consumption is equivalent to that of approximately 4,150 U.S. households \cite{de2023growing}. Globally, Schneider's report estimates that AI currently consumes 4.5 GW of power, with projections indicating a compound annual growth rate (CAGR) of 25\% to 33\% \cite{avelar2023ai}. By 2028, AI-related power consumption could reach 14 GW to 18.7 GW. This rapid growth in energy consumption poses significant challenges not only for individual data centers but also for national and global power infrastructures \cite{guo2024bridge}. 

Moreover, managing large-scale AI inference systems introduce new complexities. These systems typically utilize heterogeneous computer architectures, comprising conventional hosts (e.g., an x86 processor) and high-throughput accelerators (e.g., H100, TPU), which are fully exploited only at large batch sizes \cite{li2022ai}. The introduction of new services by cloud platforms and enterprises, such as serverless architectures, adds to the complexity of managing AI workloads effectively \cite{bianchini2024datacenter}. These factors, combined with the increasing frequency and scale of AI-induced transients, underscore the need for a shift of focus from mere power consumption to the broader, more disruptive impacts that AI could have on power grid transient dynamics and stability~\cite{zhang2018load}.

Achieving ubiquitous access to LLMs and GenAI will require timely commitments to tailored system-level planning and reliable operation of electrical grids \cite{radovanovic2021power, he2024long}. Without these commitments, the rapid growth of AI could lead to significant disruptions in power system reliability and sustainability. While the current focus on AI's electricity consumption is important, it may be overlooking a more critical aspect: the unique and potentially destabilizing effects AI workloads can pose on power grid dynamics.

\begin{figure*}[htbp]
  \centering
  \includegraphics[width=0.8\textwidth]{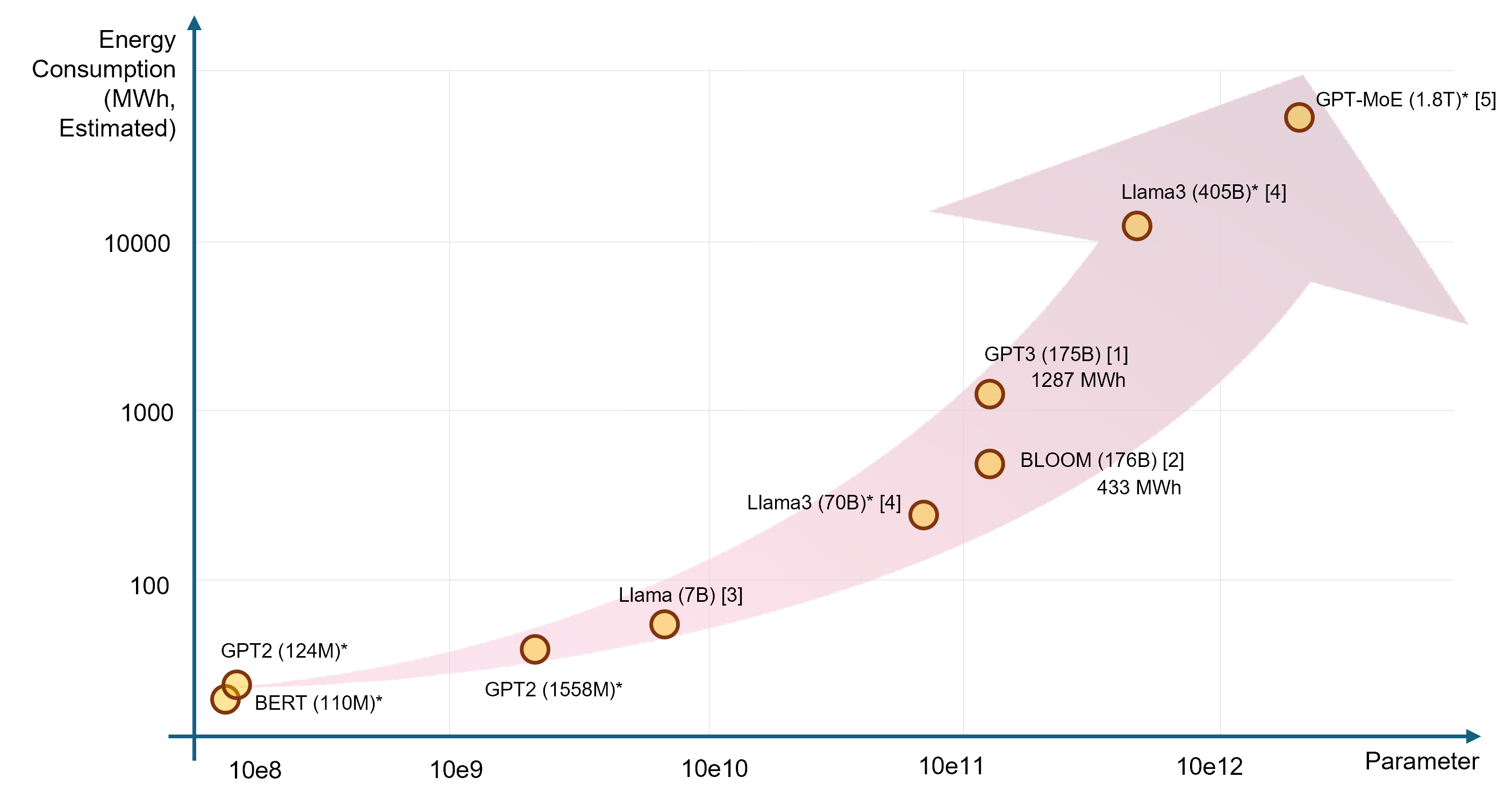}
  \caption{\small Reported energy consumption of training different LLM models with respect to model parameters \cite{de2023growing, luccioni2023estimating, mlenergy2024leaderboard, llama2024herd, elmeleegy2024demystifying}. Note the consumption shown here is relatively positioned, not based on accurate numerical calculation. The exact energy consumption can differ dramatically given different AI acceleration hardware, training and inference settings. * means estimated energy consumption based on model size. }
  \label{fig:Model_Power}
\end{figure*}

\subsection{Related Works}
Recently, emerging studies began to address the unique challenges of AI workloads by detailing the behavior of scaled-up LLMs during training and inference \cite{mcdonald2022great,10363447, patel2024characterizing, hu2024characterization, karimi2024profiling}. However, the power dynamics, particularly the transient behavior of AI facilities, remain underexplored from the power systems perspective. 

For datacenter's power behaviour, previous works have used sun irradiance and other meteorological data (i.e. temperature and wind speed) for estimating the renewable generation and renewable energy utilization  (REU) of data center load, while such approaches are rough estimates of grid-side carbon intensity \cite{ORO2015429, KWON2020115424, kong2014survey}. Some also focused on predicting the machine learning model's carbon emissions~\cite{anthony2020carbontracker}. While there are discussions on the social and environmental impacts of training and deploying AI models~\cite{kaack2022aligning}, the exact and fine-granular impacts on power grids are largely underexplored. 

As for the power performance of the AI workload, most of the work - even if they are concerned with the power consumption of AI - are more focused on either algorithmic features, steady state or quantity of energy the AI infrastructure  consumed \cite{alavani2023program, cao2023gpu}. The main focus are majorly centered around the operational or embodied electricity usage~\cite{wu2022sustainable}, the energy and natural resources that the AI models need, or the CO2 emission they may induce \cite{lacoste2019quantifying, henderson2020towards}. Pioneering studies \cite{patel2024characterizing, stojkovic2024dynamollm} characterized the power usage pattern of LLM load from a computing perspective by looking into the effects of power throttling, GPU frequency locking and etc. However, a comprehensive understanding of the transient power behavior of AI systems is still lacking,  which highlights the importance of further research in this area. Moreover, the impacts brought by AI/ML's tasks to the power grids are not fully understood nor investigated.

\subsection{Objectives and Scope of the Paper}

This paper aims to highlight the unique features of AI loads from a power and energy system perspective, and provide insights into their potential impact and solutions in light of the rapid growth of AI computing. We begin with a technical overview of modern large-scale AI models, such as LLMs, and their power consumption behavior. A holistic  mathematical description of AI-centric infrastructure is presented, followed by several case studies covering three stages of AI workload operation: training, fine-tuning, and inference. Finally, we summarize future research directions from different perspectives, including the AI-user side, data center side, and grid side, to address the various challenges ahead.

%% file: Characteristics.tex
As mentioned before, contemporary AI models, especially foundational models like LLMs, induce a great amount of energy demand, yet this is not the only concern. From an electrical power engineering perspective, the burst of such high energy demand (e.g., power surge/dip) could easily stress the local distribution system and induce more severe problems (e.g., over voltage or voltage sag on distribution line or feeders, current fluctuations, stability issues, etc.). To provide a holistic picture of how AI could impact power grids, in this section, we firstly provide qualitative analysis and the deep dive of some unique features of AI loads.

\subsection{Structure of an AI Compute Node}

\begin{figure}[htbp]
  \centering
  \includegraphics[width=0.8\columnwidth]{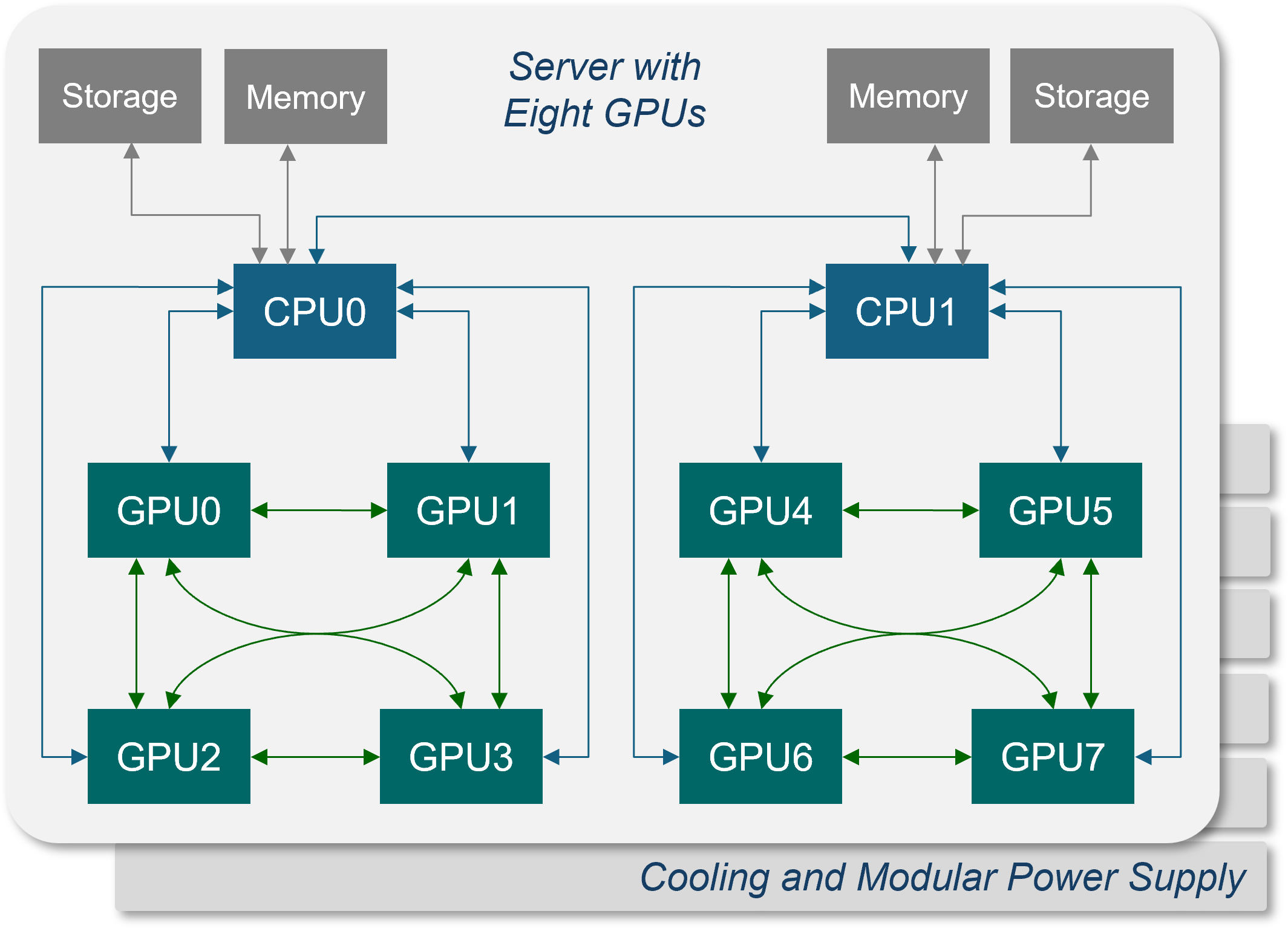}
  \caption{\small The schematic topology of an AI server with 8 GPUs.}
  \label{fig:HPC}
\end{figure}

As decipated in Figure \ref{fig:HPC}, a high-performance AI compute node typically features multiple powerful GPUs as its computational core, interconnected for efficient parallel processing. These are managed by high-core-count CPUs and supported by large amounts of high-bandwidth memory. The system incorporates high-capacity, fast storage solutions and multiple high-speed network interfaces for robust data handling and communication. Cooling is critical, with advanced systems managing the substantial heat output from densely packed components. Power delivery is designed for efficiency and redundancy, often using multiple power supply units. The architecture balances GPUs, CPUs, memory, and storage to optimize data flow and computational tasks. This design enables the node to handle intensive AI workloads, from model training to complex inferencing, while maintaining efficiency and reliability under high loads. Such systems are typically housed in large form-factor chassis, reflecting their substantial computing power and cooling requirements. 

The following table details the typical structure of a high-performance AI compute node.
\begin{table}[h]
\centering
\caption{Typical components of a high-performance AI compute node and their power considerations}
\begin{tabular}{|l|p{6cm}|}
\hline
\textbf{Component} & \textbf{Power Consideration} \\
\hline
GPU  & Primary power consumer, up to 400W-700W per GPU \\
\hline
CPU  & Moderate power draw, 200W-400W for high-end models \\
\hline
Memory  & Considerable power draw, especially for HBM (30W-80W per GPU) \\
\hline
Cooling  & Can consume 10-20\% of total system power \\
\hline
Power Supply  & Efficiency crucial, typically 90-96\% efficient or even higher \\
\hline
\end{tabular}
\end{table}

\subsection{Classification of AI Electricity Load} \label{sec:class}
Here we provide the classification and description of different operation stages for AI loads, especially LLMs.
\subsubsection{Training}
Training is the most power-intensive phase of LLM operations. During this process, transformer-based model learns from vast datasets~\cite{devlin2018bert}. Such procedure requires sustained high GPU utilization for extended periods, often lasting days, weeks, or even months. This phase demands peak power from all system components, including GPUs, CPUs, memory, and storage. The continuous high computational load generates significant heat, necessitating robust cooling systems. Power supplies must be capable of handling prolonged maximum loads, and uninterrupted power is crucial. Power capping strategies may be employed to balance performance and energy costs.

\subsubsection{Fine-tuning}
Fine-tuning involves adapting a pre-trained model to specific tasks or domains, typically requiring moderate to high power consumption. This phase sees intermittent high GPU utilization but usually for shorter durations than full training. Power draw can fluctuate more during fine-tuning, demanding cooling systems that efficiently handle variable heat loads. This phase presents opportunities for implementing power-saving features during less intensive periods and allows for more flexible power management strategies compared to full training.

\subsubsection{Inference}
Inference is generally the least power-intensive phase, involving the application of trained models to new data. It often consists of shorter computational bursts, with power consumption varying widely based on model size and query complexity. Inference workloads may benefit from specialized hardware optimized for lower power consumption. Power draw during inference can be sporadic, requiring responsive power delivery systems. This phase offers significant opportunities for power savings through efficient scheduling and hardware utilization, though it is crucial to balance low-latency response times with energy efficiency. While generally less power-intensive than training or fine-tuning, inference presents unique challenges due to its behavior-dominated nature. Usage patterns can be highly variable and unpredictable, driven by user interactions, time of day, or external events. This variability leads to rapid fluctuations in power demand, from near-idle to sudden spikes of intense computation. Moreover, user-specific application such as retrieval-augmented generation (RAG) exhibits very diverse power usage patterns. The behavior-dominated aspect of inference adds complexity to power management strategies. Systems must be capable of rapidly scaling power delivery up and down to meet changing demands while maintaining low service latency.

\subsection{The Correlation of LLM and its Hardware}

LLMs demonstrate a strong correlation between their architecture and the hardware they run on. This relationship is evident in the computational pipeline of Transformer-based LLM on a typical PC with GPU acceleration (see in Figure \ref{fig:LLM-pipline}):

\begin{figure*}[htbp]
  \centering
  \includegraphics[width=\textwidth]{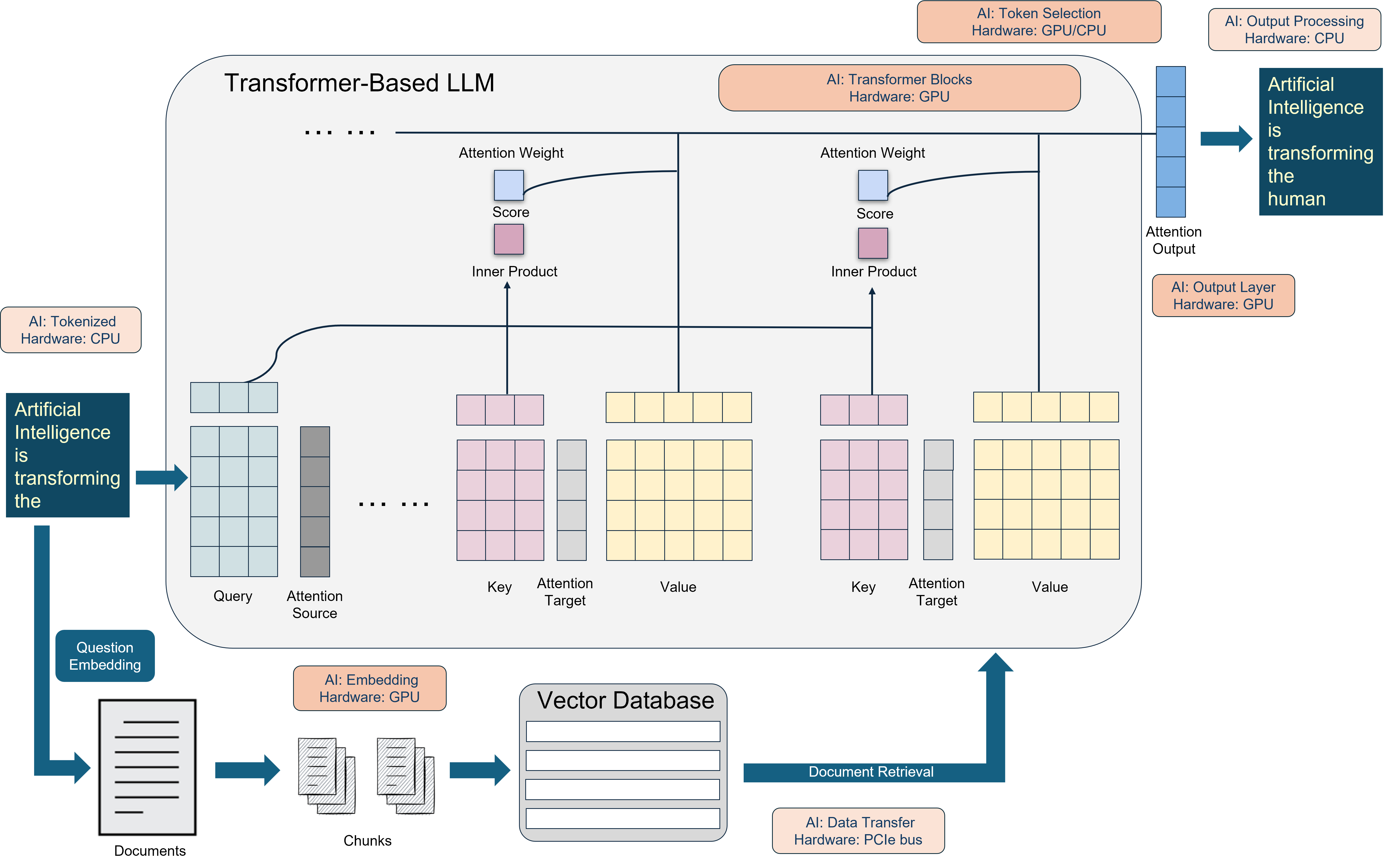}
  \caption{\small The relationship between typical LLM pipelines and associated hardware processes.}
  \label{fig:LLM-pipline}
\end{figure*}

\begin{enumerate}
    \item \textbf{Input Processing:} The CPU handles the initial tokenization of input text, a relatively light task.
    
    \item \textbf{Data Transfer:} Tokenized data moves from RAM to GPU memory via the PCIe bus, potentially becoming a bottleneck for large inputs.
    
    \item \textbf{Embedding:} The GPU's CUDA cores (assumed Nvidia) convert tokens to vector representations and add positional encodings, a moderately intensive computation.
    
    \item \textbf{Transformer Blocks:} The core of LLM's computation occurs here, leveraging the GPU's parallel processing capabilities:
    \begin{itemize}
        \item Multi-head self-attention;
        \item Layer normalization;
        \item Feed-forward neural network (MLP).
    \end{itemize}
    This stage is the most computationally demanding, showcasing the necessity of GPU acceleration for LLMs.
    
    \item \textbf{Output Layer:} The GPU performs a final linear transformation and softmax operation to produce token probabilities.
    
    \item \textbf{Token Selection:} The next token is chosen based on output probabilities, utilizing both GPU and CPU.
    
    \item \textbf{Output Processing:} The CPU converts the selected token back to text.
\end{enumerate}

For multi-token generation, steps 2-7 repeat iteratively, using each generated token as new input.

The power and computational demands of LLMs correlate strongly with model size and the running hardware. For example, the smallest version of GPT2, called GPT-2 small (124M parameters), can run on mid-range GPUs, while larger variants require high-end or multiple GPUs. (A detailed specification for typical small-scale LLM, often called small language model (SLM) can be found in Appendix.B.)  This correlation between LLMs and hardware underscores why GPUs are crucial for the efficient operation of transformer models. The GPU's capacity for parallel computation is essential for the numerous matrix operations in Transformer blocks, forming the computational backbone of modern LLMs.

\subsection{Unique features of AI workloads}

%\textcolor{red}{Most of LLM operators are compute bound, while some others are memory bound.}

\begin{figure}[htbp]
  \centering
  \includegraphics[width=\columnwidth]{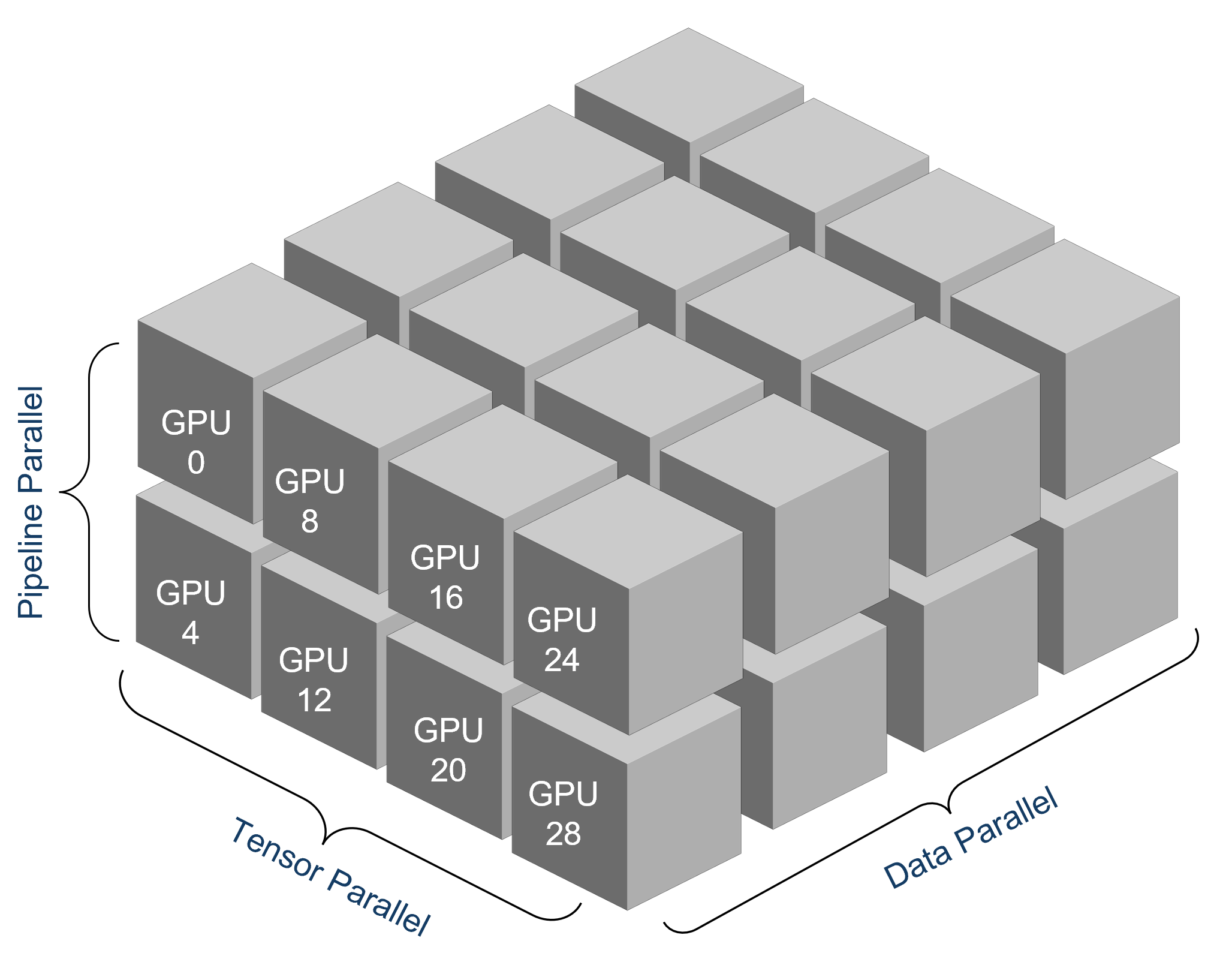}
  \caption{\small Different parallelization paradigms for AI/ML tasks~\cite{jia2019beyond}. }
  \label{fig:AI-parallel}
\end{figure}

\subsubsection{High Computational Intensity}
AI tasks, especially during training, require massive amounts of floating-point operations. For instance, training a state-of-the-art natural language processing model can involve more than $10^{8}$ PetaFLOPS (FLOPS: floating-point operations per second). This level of computational intensity is orders of magnitude higher than some traditional high-performance computing tasks. This leads to a high energy consumption over different stages of the AI lifecycle.

\subsubsection{Variability and Unpredictability}
The power consumption of AI workloads can fluctuate rapidly based on the stage of computation and the nature of the data being processed. This variability can occur on timescales of milliseconds to hours, even days. For example, during the training of deep neural networks, the backpropagation phase typically consumes more power than the forward pass, leading to periodic spikes in energy demand. Another typical example is the inference of LLMs, which can exhibit quite dramatic power peaks due to user behaviour. The power patterns are multifaceted and determined by a mixture of various factors, and sometimes unpredicatable, e.g., multiple unplanned shut downs happened during the 54-day long training process of the recent released Llama 3 405b \cite{llama2024herd}.

\subsubsection{Scalability and Non-linear Scaling}
AI deployments range from edge devices with constrained resources to massive data center clusters, each presenting unique power management challenges. The power consumption can scale from watts to megawatts. Moreover, the computational requirements and associated power consumption of AI models often scale super-linearly with model size, leading to exponential increases in energy demands as models grow larger \cite{}. However, this is assuming the physical constraints are relatively relieved. In practice, the AI accelerator normally features non-linearity and the semiconductor is also experiencing rapid development (with higher efficiency each generation) which adds another layer of complexity when estimating the power consumption.

\subsubsection{Algorithmic Sensitivity}
Small changes in AI algorithms or hyperparameters can lead to significant variations in computational requirements and, consequently, power consumption. For example, changing the learning rate or batch size during training can dramatically affect the convergence time and energy usage. Readers can refer to the Appendix.A where we demonstrate how the fine-tune algorithm relates with different power consumption stages.

\subsubsection{24/7 Operation}
Many AI systems, especially those deployed in cloud services or continuous learning scenarios, operate around-the-clock. For instance, LLM training needs to be continuously monitored. 
Such constant, uninterruptible operation leads to sustained high power demand requiring consistent power quality. Such requirement can be also contrastive to typical IT loads which may have either more pronounced diurnal patterns or more flexibility.

Table \ref{tab:ai_vs_traditional} provides a comparison between AI workloads and traditional computing loads.

\begin{table}[!t]
\renewcommand{\arraystretch}{1.3}
\caption{AI vs. Traditional Computing Loads}
\label{tab:ai_vs_traditional}
\centering
\begin{tabular}{|C{2cm}|C{2.5cm}|C{2.6cm}|}
\hline
\textbf{Aspect} & \textbf{Traditional} & \textbf{AI Workloads} \\
\hline
Power Density & 10-15 kW/rack & 30+ kW/rack \\
\hline
Load Variability & Stable, predictable & Highly variable, bursty \\
\hline
Computational Intensity & Moderate & Extremely high \\
\hline
Data Dependency & Moderate & High \\
\hline
Phase Change & Rare & Training/Fine-tuning/Inference \\
\hline
\end{tabular}
\end{table}

%% file: Modeling.tex
%To effectively manage and optimize AI infrastructure as a load on power grids, it is crucial to develop accurate mathematical models that capture the unique characteristics of AI workloads. This section presents a comprehensive mathematical framework for characterizing the power consumption patterns of AI systems. It highlights the loads of training phases and the dynamic loads of inference operations. 

Due to the aforementioned features of the AI loads, the detailed analytical/deterministic model of AI is quite difficult to build, and sometimes, impossible to acquire. However, this does not mean the mathematical description is not beneficial. In contrast, to effectively manage and optimize AI infrastructure as a load on power grids, it is crucial to develop mathematical models as detailed and accurate as we can to achieve proactive operation of local power grids with high penetration of AI infrastructure. Following this idea, this section addresses the AI-centric data center models with high-level mathematical language.

%\subsection{Power Measurements}
\subsection{Metrics for AI-centric Data Center}
Several power measurements come naturally into our view when considering the dynamics of AI loads. These metrics provide crucial insights into power consumption patterns, efficiency, and potential challenges in managing AI workloads in data centers. 

\subsubsection{TDP (Thermal Design Power)}

TDP, or thermal design power, is defined as the maximum power that a subsystem is allowed to draw for a "real world" application. TDP is usually determined by the needs of the component that needs to be cooled. In the context of AI workloads, TDP is critical for cooling system design in data centers. AI tasks, especially training large models, often push hardware to its TDP limits. Understanding TDP helps in capacity planning and ensuring adequate cooling for AI hardware, which is essential for maintaining optimal performance and longevity of the equipment.

\subsubsection{GPU Utilization}

GPU utilization, while not strictly a power measurement, is closely related to power consumption in AI workloads. It represents the percentage of GPU computational resources being used. High GPU utilization is typical for AI training tasks and correlates strongly with high power consumption. Monitoring GPU utilization helps in optimizing workload distribution and energy efficiency, and can be used to identify opportunities for workload consolidation or expansion.

\subsubsection{PUE (Power Usage Effectiveness)}

PUE, or Power Usage Effectiveness, is the ratio of total energy used by a data center to the energy delivered to computing equipment. AI workloads can significantly impact PUE due to their high computational demands. Optimizing PUE is crucial for energy-efficient AI operations, and AI-centric data centers may need innovative cooling solutions to maintain good PUE. As AI workloads become more prevalent, maintaining an efficient PUE becomes increasingly challenging and important for sustainable data center operations.

\subsubsection{Peak/Average Ratio}

The Peak/Average ratio is the ratio of peak power consumption to average power consumption. This metric is particularly interesting for AI-centric data centers under long-term training jobs, which show a Peak/Average ratio close to 1, compared to 1.5-2.0 for conventional data centers. This indicates that AI workloads maintain consistently high power consumption during operation. Such a characteristic has significant implications for power supply design and energy management in AI facilities, necessitating robust and stable power delivery systems.

\subsubsection{Peak/Idle Ratio}

The Peak/Idle ratio, often overlooked in current literature, is the ratio of peak power consumption to idle power consumption. This metric is crucial for understanding power transience in AI workloads. It indicates how significant the power drop is after training interruption and how large the AI load could change the local power flow. High Peak/Idle ratios in AI workloads can cause substantial local power flow changes, making it an important consideration for designing power systems that can handle rapid transitions between active and idle states.

\subsubsection{dP/dt (Rate of Change of Power)}

The dP/dt metric, representing the rate at which power consumption changes over time, is also often overlooked in current work. For a mega-watt scale AI data center, the unplanned stop of training can cause internal power disruption within few seconds or even shorter, resulting in great transience. If this cannot be buffered by the redundant Energy Storage System (ESS), then it will affect the local grid. This metric is critical for understanding and managing the dynamic nature of AI workloads and their potential impact on power infrastructure.

While dP/dt provides a general measure of power change over time, the ramping rate and decline rate offer more specific insights into the behavior of AI systems during periods of increasing and decreasing power consumption, respectively.

The ramping rate represents the speed at which power consumption increases, typically observed during the startup of AI workloads or transitions to higher computational intensities. Conversely, the decline rate describes the speed of power reduction, often seen during job completion, sudden terminations, or transitions to lower computational states.

These rates are particularly important for characterizing the transient behavior of AI workloads. They play a crucial role in:

\begin{itemize}
    \item Designing power delivery systems capable of handling rapid fluctuations;
    \item Optimizing job scheduling to minimize stress on power infrastructure;
    \item Sizing and configuring energy storage systems to buffer transients;
    \item Ensuring grid stability in the presence of large AI computing facilities.
\end{itemize}

In the context of AI workloads, both ramping and decline rates can be exceptionally high. For instance, the initiation of a large-scale training job might cause a near-instantaneous surge in power demand as GPUs are fully engaged. Similarly, the completion or interruption of such a job can lead to a rapid decline in power consumption.

These metrics and their mathematical representations will be central to our subsequent analysis of AI power dynamics. We will develop models that capture the unique characteristics of AI workloads, such as sustained high power draw, rapid fluctuations, and the relationship between computational load and power consumption. To ease the modelling and maintain a certain universality, the data center example from Google is selected as our benchmark system. Please note that the following modelling methodology can also be applied for other AI-centric data center configurations with minimum tuning on certain parameters like the distribution system type (DC bus, or AC bus), power station type (AC/AC or AC/DC), different supporting infrastructure, different external energy sources, etc.

\begin{figure*}[htbp]
  \centering
  \includegraphics[width=0.9\textwidth]{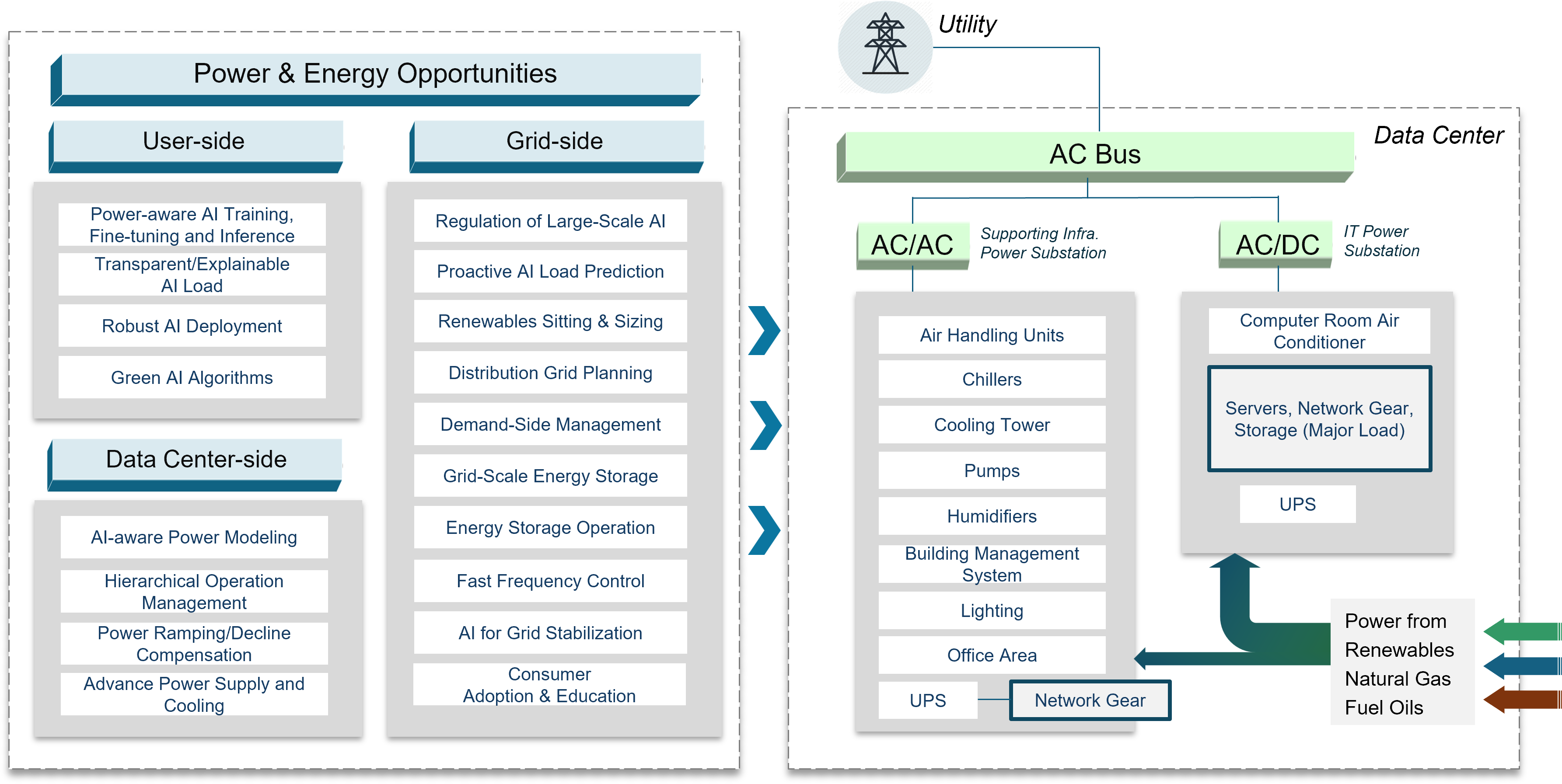}
  \caption{\small Details of Data Center Electrical System (an example from Google \cite{google2024datacenters}).}
  \label{fig:Model-System}
\end{figure*}

\subsection{Total Data Center Power Consumption}
To analyze the data center's power usage, one aspect is to look into the total power consumption of the data center, which can be modeled as:

\begin{equation}
\label{equ:P_total}
P_{\text{total}} = P_{\text{AC\_Bus}} + P_{\text{External}}.
\end{equation}

In \eqref{equ:P_total}, $P_{\text{AC\_Bus}}$ denotes the power supplied by the power utility from distribution grids, and $P_{\text{External}}$ represents all external  power and energy sources, such as renewable energy generation, natural gas, fuel oils, etc. Recently, researchers and engineers have also explored the potential of hydrogen as an alternative power source~\cite{mcqueen2020department}.

The total power is distributed between the Supporting Infrastructure and IT Power Substation:

\begin{equation}
P_{\text{total}} = P_{\text{Supporting\_Infra}} + P_{\text{IT\_Power}}.
\end{equation}

The power consumption of the supporting infrastructure can be modeled with the major components:

\begin{equation}
\begin{split}
P_{\text{Supporting\_Infra}} = & \eta_{\text{AC/AC}} \cdot (P_{\text{AHU}} + \\
&P_{\text{Chillers}} + P_{\text{CoolingTower}} + P_{\text{Pumps}} + \\
&P_{\text{Humidifiers}} + P_{\text{BMS}} + P_{\text{Lighting}} + \\
&P_{\text{Office}} + P_{\text{UPS\_Infra}} + P_{\text{Network\_Infra}});
\end{split}
\end{equation}
Where $\eta_{\text{AC/AC}}$ represents the efficiency of the AC/AC conversion, AHU denotes the air handling unit, BMS is the building management system.

The IT power consumption, including the AI workload, is made up of the following major components:

\begin{equation}
\begin{split}
P_{\text{IT\_Power}} = &\eta_{\text{AC/DC}} \cdot (P_{\text{Servers}} + P_{\text{NetworkGear}} + \\
&P_{\text{Storage}} + P_{\text{CRAC}} + P_{\text{UPS\_IT}});
\end{split}
\end{equation}
Where $\eta_{\text{AC/DC}}$ represents the efficiency of the AC/DC conversion, CRAC is computer room air conditioner. (Here we omit some details like the power distribution units (PDUs), switches, transformers, circuit breakers, etc.)
Traditionally, a data center can also use an AC/AC conversion here, which normally involves several power conversion stages. With an AC/DC conversion, the internal local grid becomes a DC system and can be considered as a DC microgrid~\cite{salomonsson2007adaptive}.

\subsection{AI Workload Power Consumption}

\subsubsection{Power Usage Breakdown of AI Accelerators}

In this work, we assume all AI computation is happening in the servers, and their power consumption can be expressed as:

\begin{equation}
P_{\text{Servers}} = P_{\text{AI}} + P_{\text{Other\_Compute}};
\end{equation}
Where $P_{\text{AI}}$ represents the AI workload power consumption in the servers, and $P_{\text{Other\_Compute}}$ denotes all the related power needs that do not involve AI loads.

If steady-state power consumption is the main focus, a simpler model can be used (assuming the same GPUs are used and computation burden is shared evenly among them):

\begin{equation}
P_{\text{AI}} \approx N \cdot P_{\text{peak}} \cdot U_{\text{phase}};
\end{equation}
Where $N$ is the number of AI accelerators, $P_{\text{peak}}$ is the peak power consumption of a single AI accelerator (this value is normally very close to TDP), $U_{\text{phase}}$ is the utilization factor (phase can be pre-train, fine-tune, or inference), which is a key metric representing the proportion of an AI accelerator's capacity being used at time t during a specific operational phase. This factor ranges from 0 (idle) to 1 (full utilization), providing crucial insights into the system's workload status. During pre-training $U_{\text{pretrain}}$, the utilization factor often approaches 1 for extended periods, indicating near-constant full utilization as the model learns from vast datasets. In the fine-tuning phase $U_{\text{finetune}}$, utilization remains high but becomes more variable, typically averaging between 0.7 to 0.9 as the model is adapted to specific tasks. The inference phase $U_{\text{inference}}$ exhibits the most variability in utilization, potentially covering the entire range from 0 to 1. This wide variation in $U_{\text{inference}}$ reflects the behavior-dominated nature of inference workloads, where utilization fluctuates based on real-time demand, user interactions, and external factors. Understanding these utilization patterns across different phases is essential for effective power management and resource allocation in AI systems.

\subsection{Dynamic Power Consumption}

Accounting for the dynamic nature of AI workloads and potential variability in external energy supply:

\begin{equation}
P_{\text{dynamic}}(t) = P_{\text{AI}}(t) + C \cdot \frac{dP_{\text{AI}}}{dt} + E \cdot \frac{dP_{\text{external}}}{dt};
\end{equation}
Where $C$ and $E$ are coefficients representing the scale of the system's dynamics on AI power changes and external power fluctuations, respectively.

For a simpler case, we can omit the external power effect and this will give us:

\begin{equation}
P_{\text{dynamic}}(t) = P_{\text{AI}}(t) + C \cdot \frac{dP_{\text{AI}}}{dt}.
\end{equation}

Building upon our previous model, we can further highlight the dynamic and transient features:
\begin{equation} 
P_\text{dynamic}(t) = P_\text{AI}(t) + C \cdot \frac{dP_\text{AI}}{dt} + D \cdot \frac{d^2P_\text{AI}}{dt^2};
\end{equation}
Where we have added a second-order term $D \cdot \frac{d^2P_\text{AI}}{dt^2}$ to capture the rapid power fluctuations. Here, $C$ represents the scale of the system's gradual power changes, while $D$ represents its scale to abrupt changes.

In an AI-centric data center, the majority of energy will be consumed by AI-specific hardware, i.e., AI accelerators (such as NVIDIA's A100 or H100 GPUs, are essentially specialized, high-performance computers). Their power consumption can be broken down into several key components:

\begin{equation}
\begin{split}
    P_\text{AI}(t) \approx P_\text{AI\_accelerator}(t) = P_\text{compute}(t) + P_\text{memory}(t) + \\
    P_\text{cooling}(t) + P_\text{auxiliary}(t);
\end{split}
\end{equation}
Where $P_\text{compute}(t)$: Power consumed by the computation units (GPU cores, tensor cores), $P_\text{memory}(t)$: Power consumed by memory operations (HBM, cache), $P_\text{cooling}(t)$: Power consumed by cooling systems (fans, liquid cooling), $P_\text{auxiliary}(t)$: Power consumed by other components within the rack (mother board, communications, voltage regulators, etc.).

We now provide a formal mathematical description of the three different operation stages for AI loads, particularly focusing on LLMs (see description of training, fine-tuning, inference in section \ref{sec:class}).

\subsubsection{Training}

Training, the most power-intensive phase, can be mathematically represented as:

\begin{equation}
P_\text{training}(t) = P_\text{base} + \sum_{i=1}^{N} P_\text{AI\_accelerator\_i}(t).
\end{equation}

The sustained high utilization can be modeled as:

\begin{equation}
U_\text{AI\_accelerator}(t) \approx U_\text{max} \quad \text{for } t \in [t_\text{start}, t_\text{end}];
\end{equation}
Where $U_\text{AI\_accelerator}(t)$ is the AI accelerator utilization at time $t$, and $[t_\text{start}, t_\text{end}]$ represents the duration of the training phase.

\subsubsection{Fine-tuning}

Fine-tuning can be represented as

\begin{equation}
    P_\text{fine-tuning}(t) = P_\text{base} + \alpha(t) \cdot \sum_{i=1}^{N} P_\text{AI\_accelerator\_i}(t);
\end{equation}
Where $\alpha(t)$ is a time-dependent scaling factor representing the variable AI accelerator utilization during fine-tuning:

\begin{equation}
\alpha(t) = \begin{cases} 
    1 & \text{during high utilization periods}; \\
    \beta & \text{during lower utilization periods, where } 0 < \beta < 1.
\end{cases}
\end{equation}

\subsubsection{Inference}

Inference, characterized by its behavior-dominated nature, can be modeled as:

\begin{equation}
P_\text{inference}(t) = P_\text{base} + \sum_{j=1}^{M} B_j(t) \cdot P_\text{query}(s_j, c_j).
\end{equation}

The behavior-dominated aspect can be captured by modeling $B_j(t)$ as a stochastic process, e.g., Poisson distribution:

\begin{equation}
B_j(t) \sim \text{Poisson}(\lambda(t)).
\end{equation}

Where $\lambda(t)$ is a time-dependent rate parameter that may vary based on factors such as time of day or external events.

The overall power consumption of an AI system can then be expressed as:

\begin{equation}
\begin{split}
    P_\text{AI}(t) = w_\text{train}(t) \cdot P_\text{training}(t) +\\
    w_\text{finetune}(t) \cdot P_\text{fine-tuning}(t) + \\
    w_\text{inference}(t) \cdot P_\text{inference}(t);
\end{split}
\end{equation}
Where $w_\text{train}(t)$, $w_\text{finetune}(t)$, and $w_\text{inference}(t)$ are time-dependent weights indicating the proportion of resources allocated to each phase at time $t$ in the data center. In a large scale AI-centric data center, three phases of AI loads can co-exist at the same time, however, for a single node or a computer with single GPU, normally only one type should be deployed.

To quantify the total energy $E_{\text{total}}$used for AI models, we can do integration of power consumption over time as $E_{\text{total}} = \int_0^T P_{\text{total}}(t) dt.$

%% file: simulation.tex
To demonstrate the potential impacts brought by AI load, in this section, we simulate and analyze the power consumption characteristics of AI models under various settings of model size, hardware, and job types. 

\subsection{Case study 1: Data center behaviour}
The MIT Supercloud Dataset was collected on the TXGaia system, a heterogeneous cluster at MIT's datacenter \cite{samsi2021supercloud}. This system consists of 224 GPU-accelerated nodes, each equipped with two 20-core Intel Xeon Gold 6248 processors (384GB RAM) and two NVIDIA Volta V100 GPUs (32GB RAM each), as well as additional CPU-only nodes. The dataset, totaling over 2.1 TB, includes time series data on CPU and GPU utilization, memory usage, GPU temperature, node state snapshots, file I/O, and scheduler logs. The dataset features labeled data from running and manually annotating common deep neural networks across vision, Natural Language Processing (NLP), and Graph Neural Networks (GNN). In detail, it includes various vision networks (VGG, ResNet, Inception, U-Net), language models (BERT, DistilBERT), and graph neural networks (DimeNet, SchNet, PNA, NNConv).

BERT (Bidirectional Encoder Representations from Transformers~\cite{devlin2018bert}) is one of the language models in this dataset, with 189 job counts recorded. BERT is a foundational transformer-based machine learning model for NLP tasks, pre-trained on a large corpus of unlabeled text. BERT is designed to understand the context of words in search queries and has significantly improved the interpretation of natural language in various applications.

\begin{figure*}[htbp]
  \centering
  \includegraphics[width=\textwidth]{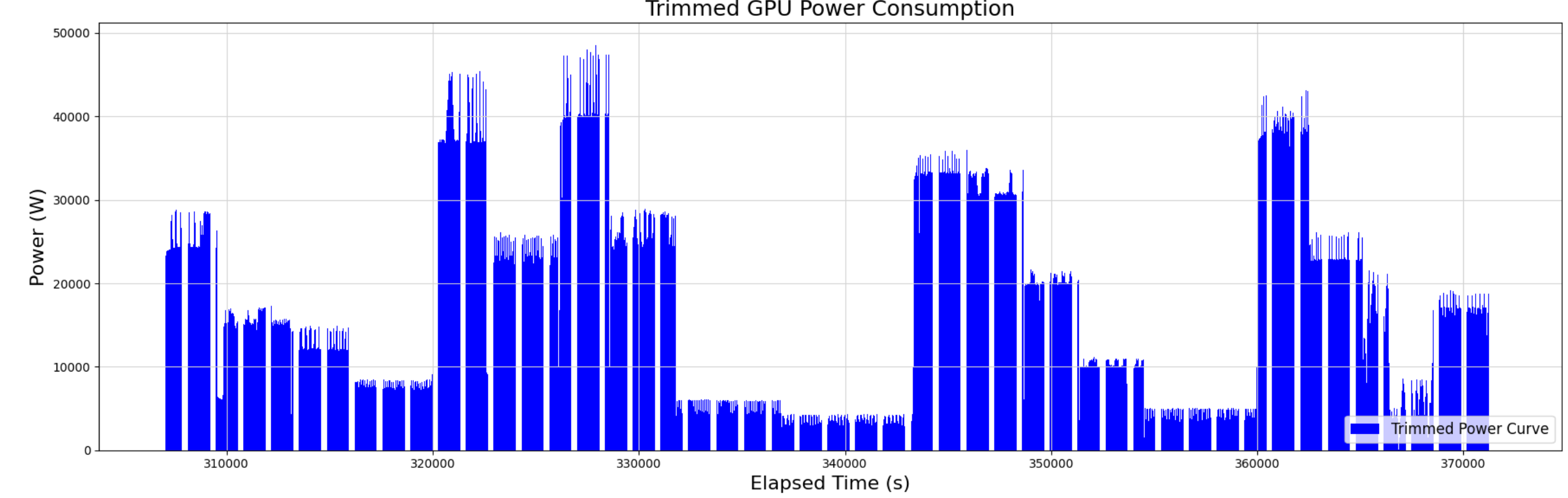}
  \caption{\small Power Consumption of BERT in MIT Supercloud Dataset (Peak power consumption of approximately 48.70 kW, with an average consumption of 17.80 kW and a standard deviation of 12.39 kW. The job ran for an extended period of 4 days).}
  \label{fig:BERT}
\end{figure*}

\begin{figure}[htbp]
  \centering
  \includegraphics[width=\columnwidth]{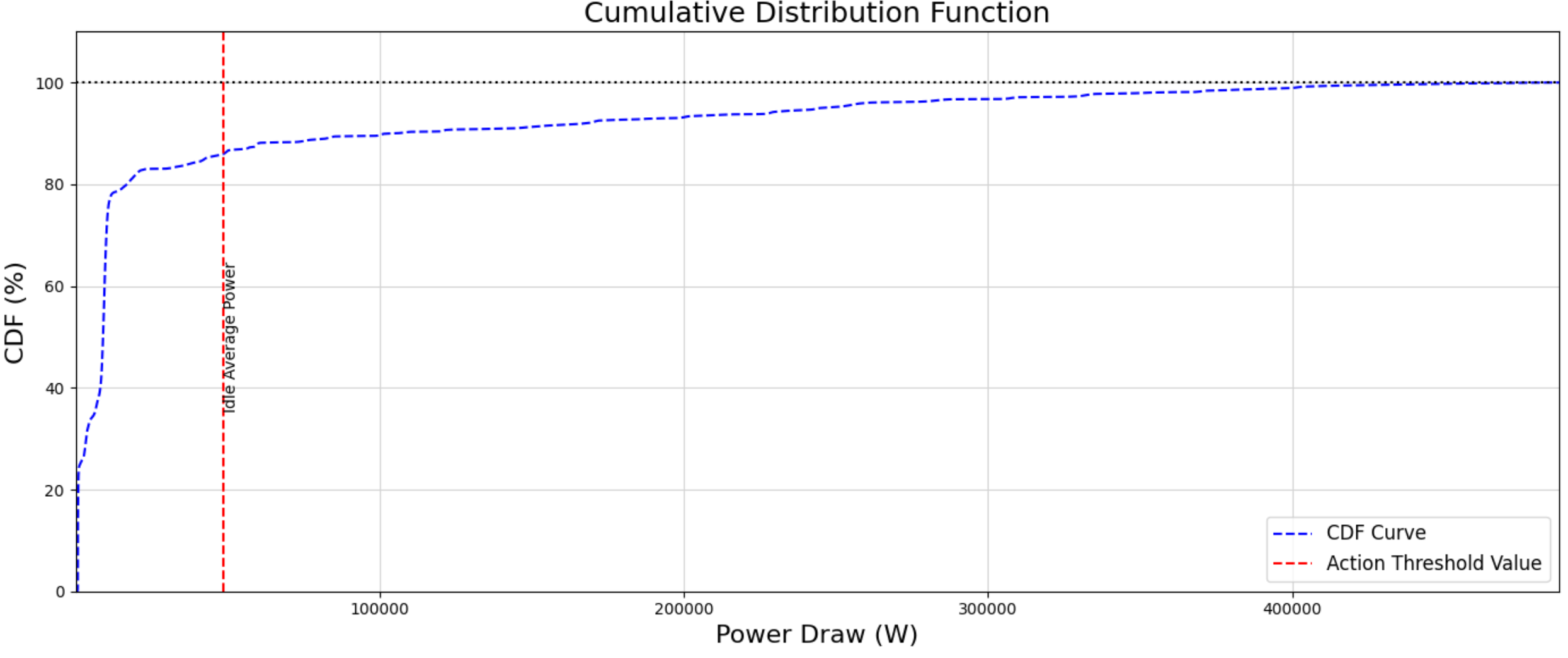}
  \caption{\small Power Consumption CDF of BERT.}
  \label{fig:BERT-CDF}
\end{figure}

\begin{figure}[htbp]
  \centering
  \includegraphics[width=\columnwidth]{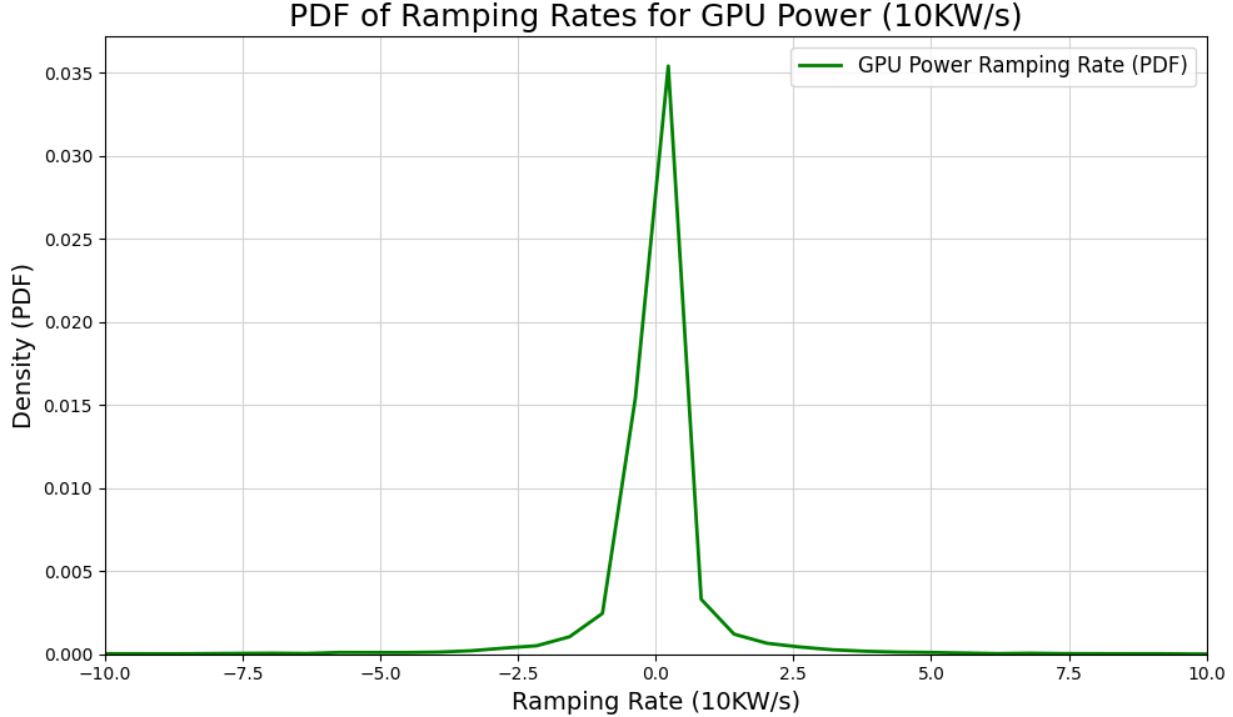}
  \caption{\small Power Consumption PDF of BERT.}
  \label{fig:BERT-PDF}
\end{figure}

The power consumption results for a specific BERT job reveal significant insights into the system's power dynamics. As the Figure \ref{fig:BERT} shows, the BERT model's power consumption is highly dynamic, with frequent transitions between low and high power states. While it can reach peaks of nearly 50kW, it often operates at much lower power levels. The system experiences both gradual and rapid changes in power consumption, likely corresponding to different phases of model operation such as data loading, forward passes, backward passes, and parameter updates. The action threshold at 50kW in the CDF graph (Figure \ref{fig:BERT-CDF}) suggests that power draws above this level may be considered unusual or require special handling. The PDF of ramping rates (Figure \ref{fig:BERT-PDF}) indicates that while the system can change power rapidly, it tends to favor more gradual transitions or stable states. This power profile is consistent with the complex and computationally intensive nature of training large language models like BERT, which involve alternating periods of high computational load and data movement or synchronization steps.

%INCLUDE OTHER PLOTS IF NEEDED

\subsection{Baselines: LLMs}

Consider the sizes of the models, the size of the data, along with the hardware limits on memory and computation, we are curating a list of results from experiments with feasible configurations as shown in Table \ref{table:desktop_setups_detailed}. Here, We present the training performance of GPT2-124M using Setup 1 and nanoGPT using Setup 2; the fine-tuning performance of GPT2-medium using Setup 2; the inference performance of nanoFPT, GPT-medium, Mamba and Transformer using Setup 2. The model details can be found in Table \ref{tab:model-specs} of Appendix.B. The GPU details can be found in Table \ref{tab:gpu-comparison} of Appendix.C.

\setlength\cellspacetoplimit{4pt}
\setlength\cellspacebottomlimit{4pt}

\begin{table}[h!]
\caption{LLM Case Study Setups, Hardware, and Tasks}
\centering
\begin{tabular}{|p{1cm}|p{2cm}|p{2cm}|p{2cm}|}
\hline
\textbf{Setup} & \textbf{Hardware} & \textbf{Task Type} & \textbf{Tasks} \\ \hline
\textbf{Setup1} & Nvidia 4090\newline 1000W PSU & Pre-train & GPT2-124M training \\ \cline{3-4}
 &  & Fine-tuning & - \\ \cline{3-4}
 &  & Inference & - \\ \hline
\textbf{Setup2} & AMD 7900XTX\newline 750W PSU & Pre-train & nanoGPT training \\ \cline{3-4}
 &  & Fine-tuning & GPT2-medium fine-tuning \\ \cline{3-4}
 &  & \multirow{3}{*}{Inference} & nanoGPT inference \\ \cline{4-4}
 &  &  & GPT-medium inference \\ \cline{4-4}
 &  &  & Mamba inference \\ \cline{4-4}
 &  &  & Transformer inference \\ \hline
\end{tabular}
\label{table:desktop_setups_detailed}
\end{table}

%SOFTWARE ENV
%Setup1 Ubuntu 22.04 CUDA 12.4
%Setup2 Ubuntu 22.04 ROCM 6.1

%Here, a table is needed:
%Desktop setup1 (Nvidia 4090), 1000W power supply
%Task: GPT2-124M training
%Desktop setup2 (AMD 7900xtx), 750W power supply
%Task: nanoGPT training, GPT2-medium fine-tuning, Mamba+transformer inference

%desktop set-up1 (Nvidia 3080) 

%Desktop setup1 (Nvidia 4090), 1000W power supply
%Task: GPT2-124M training
%Intel 9900, 64GB RAM, 2T SSD

%Desktop setup2 (AMD 7900xtx), 750W power supply
%Task: nanoGPT training, GPT2-medium fine-tuning, Mamba+transformer inference
%Intel Xeon 2650, 64GB RAM, 1T SSD

%We can also include a GPU Cluster (or maybe in future), it has (9*Nvidia A40)

\subsection{Case study 2: LLM training}
%GPT2-124M pre-training on 4090, following \cite{karpathy2024}
%A compact version of GPT-2 is trained on latest 4090 GPU, 1000W PSU and power dynamics are monitored for approximately 22 hrs, figure \ref{fig:GPT2-Training} represents only high power consumption time stamp. Maximum power consumption is  461 W with average power of 414 W and standard deviation is 113.7 W.

This training section demonstrates the power consumption patterns of two different GPU setups while training language models. The models being trained are GPT-2 124M and nanoGPT, both of which are transformer-based language models, but with different architectures and sizes.

GPT-2 124M is a compact version of the GPT-2 model developed by OpenAI \cite{huggingface2024gpt2}. It contains 124 million parameters, making it the smallest in the GPT-2 family but still capable of generating coherent text and performing various language tasks. NanoGPT, on the other hand, is a minimalist implementation of the GPT architecture \cite{karpathy2024nanogpt}, designed by Andrej Karpathy to be more accessible and easier to understand than larger, more complex models.

The power consumption characteristics of these training sessions reveal interesting differences between the NVIDIA RTX 4090 and AMD Radeon RX 7900 XTX GPUs, as well as between the two model architectures. The GPT-2 124M training on the RTX 4090 shows a relatively stable power draw, averaging 414W with a maximum of 461W over a 22-hour training period. This setup exhibits large power transients, with drops of about 320W and ramps of 350W, demonstrating the GPU's ability to quickly adjust its power state.

In contrast, the nanoGPT training on the RX 7900 XTX displays more variable power consumption, fluctuating between approximately 50W and 250W. The power transients for this setup are smaller, with drops of about 150W and ramps of 130W. This difference in power profile could be attributed to the smaller model size of nanoGPT, differences in GPU architecture, or varying power management strategies between NVIDIA and AMD.

The unique power consumption patterns observed in these training sessions highlight the dynamic nature of GPU utilization in AI workloads. The RTX 4090's higher and more stable power draw, coupled with larger transients, suggests the device operating closer to its maximum capacity for longer periods. This is supported by its high average power of 414W and standard deviation of 113.7W. The RX 7900 XTX, while showing lower overall power consumption, demonstrates more frequent fluctuations, indicating algorithmic influence of the AI training coupling with the characteristics of the smaller nanoGPT model.

These observations underscore the importance of considering both hardware capabilities and model architecture when setting up AI training environments. The choice of GPU and model can significantly impact power consumption patterns, which in turn affects energy efficiency, cooling requirements, and overall system design for AI training setups.

\begin{figure*}[htbp]
  \centering
  \includegraphics[width=0.95\textwidth]{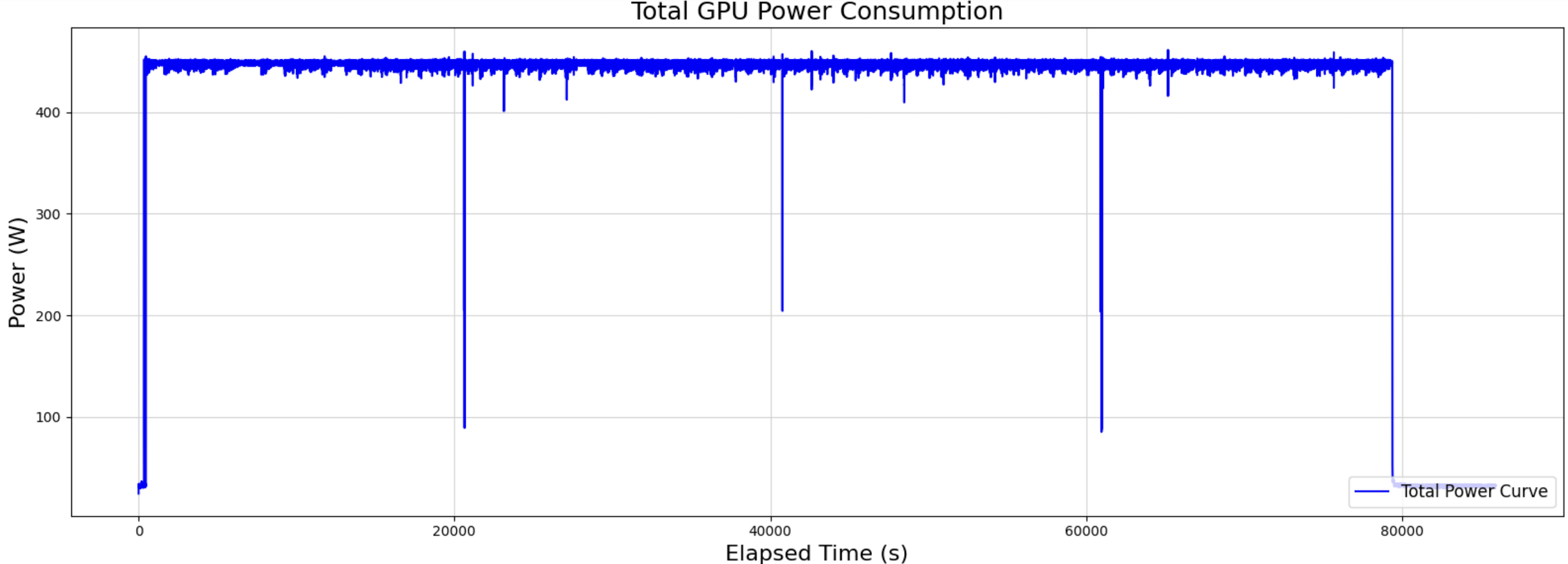}
  \caption{\small Power Consumption of GPT-2 124M trained on setup 1.}
  \label{fig:GPT2-Training}
\end{figure*}

\begin{figure*}[htbp]
  \centering
  \includegraphics[width=0.95\textwidth]{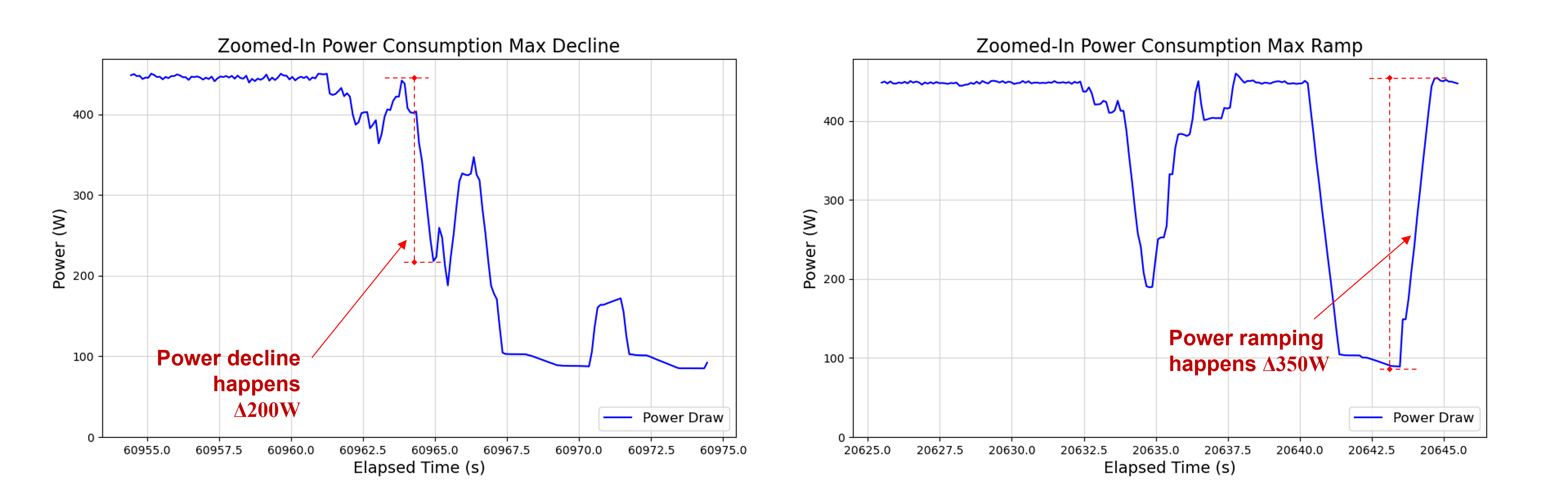}
  \caption{\small Power Transients of GPT-2 124M trained on setup 1.}
  \label{fig:GPT2-Training-Transient}
\end{figure*}

%Pre-train of nanoGPT is pre-trained on AMD 7900xtx, 750W PSU setup 2, following guideline from \cite{jia2024guide}.
% Fine tuning operation of nano GPT consumes lower amount of energy comparatively to other models with Peak power of 280W and standard deviation of 27 W approximately.

\begin{figure*}[htbp]
  \centering
  \includegraphics[width=0.95\textwidth]{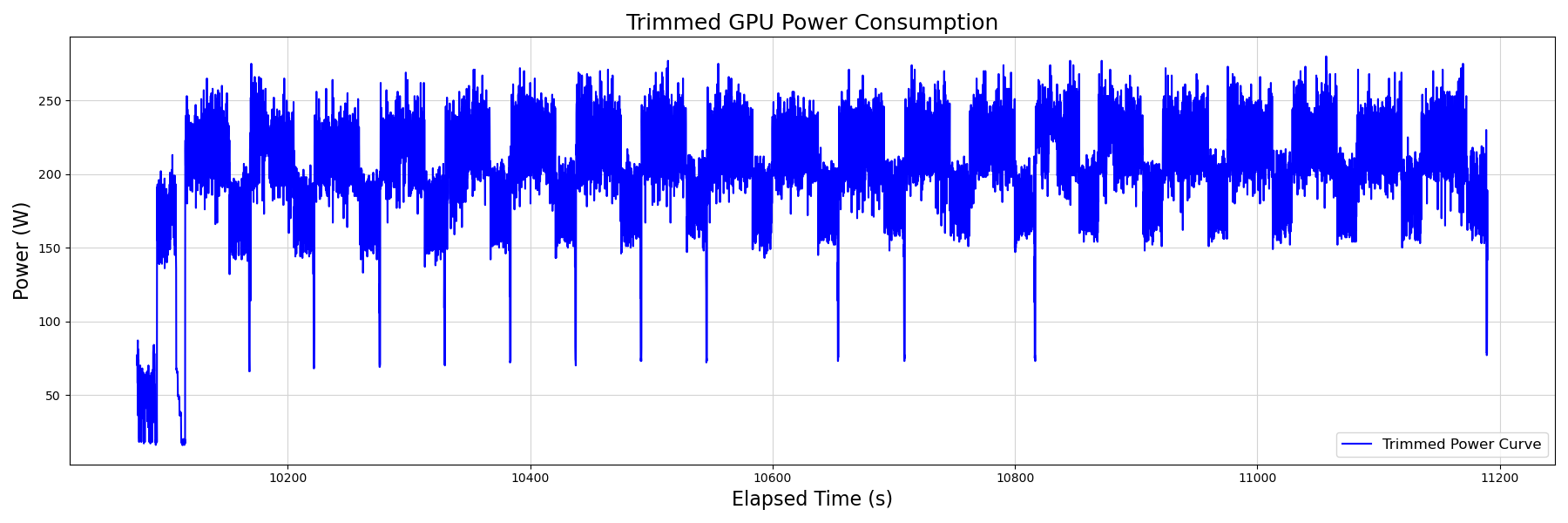}
  \caption{\small Power Consumption of nanoGPT trained on setup 2.}
  \label{fig:nanoGPT-Training}
\end{figure*}

\begin{figure*}[htbp]
  \centering
  \includegraphics[width=0.95\textwidth]{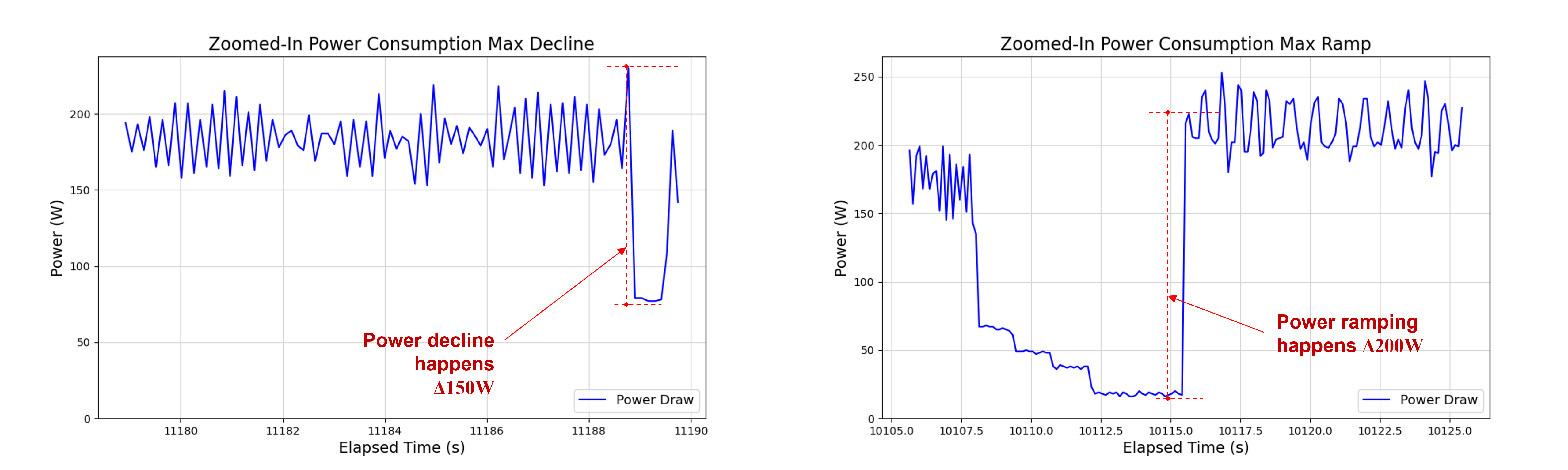}
  \caption{\small Power Transients of nanoGPT trained on setup 2.}
  \label{fig:nanoGPT-Training-Transient}
\end{figure*}

\subsection{Case study 3: Fintuning of LLM}

\begin{figure*}[htbp]
  \centering
  \includegraphics[width=0.95\textwidth]{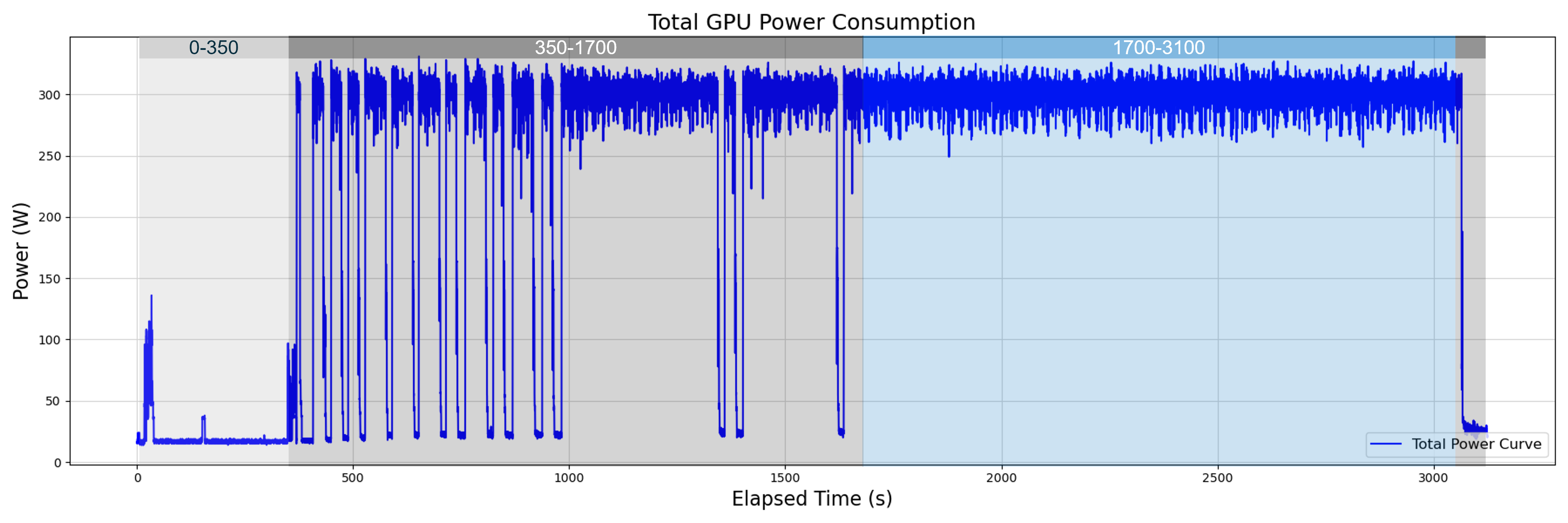}
  \caption{\small Power Consumption of GPT2-medium fine-tuned on AMD GPU 7900 XTX.}
  \label{fig:GPT2-medium-finetune}
\end{figure*}

%GPT2-medium fine-tune on 7900xtx following guideline from \cite{jia2024guide}
The fine-tuning process of GPT2-medium on an AMD GPU 7900 XTX, following AMD's guidance \cite{jia2024guide}, demonstrates distinct power consumption patterns that correspond to different stages of the training process. The power consumption graph illustrates the dynamic nature of GPU utilization during model fine-tuning, with four key phases identifiable as shown in Figure \ref{fig:GPT2-medium-finetune}.

The initialization and setup phase (0-350s) shows an initial spike in power consumption followed by a period of low power draw, reflecting the model initialization, data loading, and optimizer setup. The early training and learning rate warmup phase (350-1700s) is characterized by high and fluctuating power consumption, with peaks reaching around 330W and regular sharp drops likely corresponding to evaluation intervals. This erratic pattern may be attributed to the learning rate warmup and aggressive parameter updates. As the training progresses into the late training and learning rate decay phase (1700-3100s), the power consumption remains high but appears slightly more consistent, potentially due to the learning rate entering its decay phase and more refined parameter updates. The training completion and shutdown phase (3100-3200s) is marked by a sharp drop in power consumption, signaling the end of the training process. A more detailed analysis can be found in Appendix.A.

Throughout the main training phases, the power consumption predominantly fluctuates between approximately 250W and 330W, with regular drops to near 0W during evaluation periods. This detailed power consumption profile provides valuable insights into the GPU's behavior during fine-tuning, highlighting the importance of agile power management in AI training setups and demonstrating how different training stages impact GPU utilization.

\subsection{Case study 4: LLM Inference}
%Power matrices in LLM inference section considers two scenarios \cite{jia2024guide}: i) GPT-2 ii) Nano GPT,  with approximate maximum power consumption during inference jobs is 300 W and elapsed time is between 25 to 50 sec. Standard deviation is around 50 W with maximum energy consumption of 1.57kJ.

The case study on LLM inference power consumption, focusing on GPT-2 and nanoGPT models, reveals intriguing patterns in GPU energy utilization. Both models demonstrate similar overall power profiles during inference, characterized by rapid transitions between low-power idle states and high-intensity computation phases. These power surges reach peaks of approximately 300W and last between 25 to 50 seconds, with a standard deviation of about 50W, indicating significant variability in power draw.

\begin{figure*}[htbp]
  \centering
  \includegraphics[width=\textwidth]{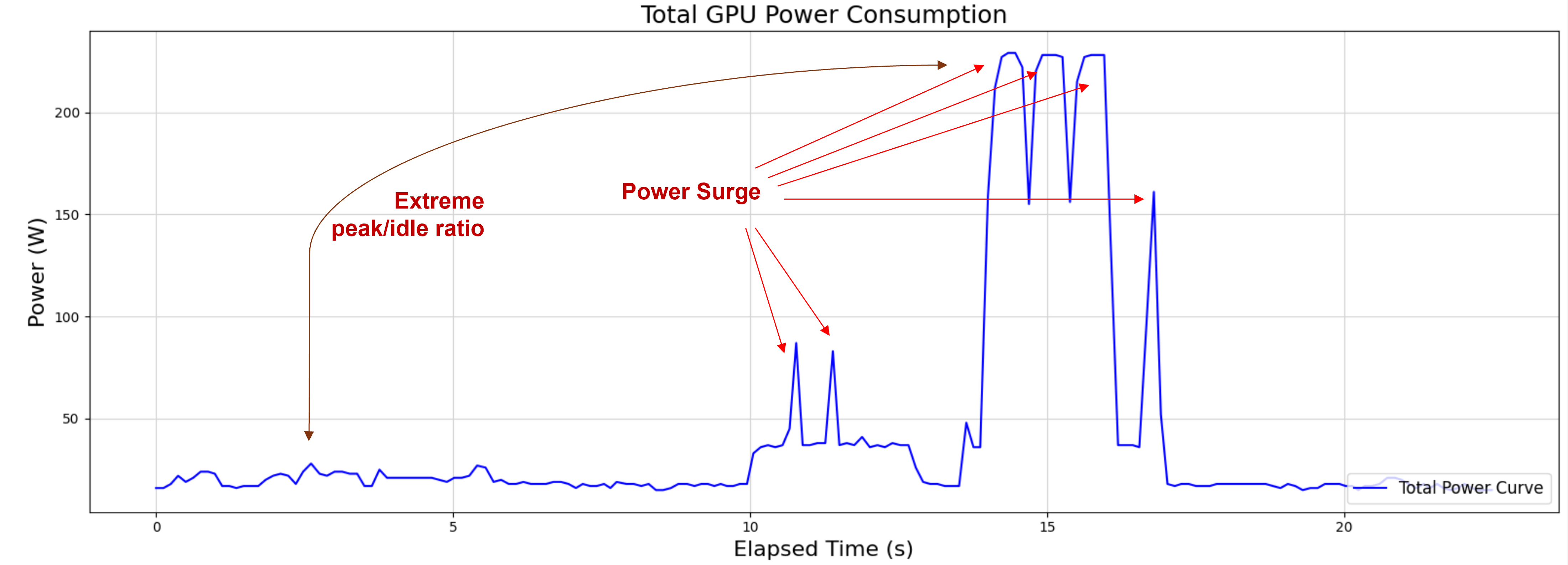}
  \caption{\small Power Consumption of Inference of nanoGPT running on setup 2.}
  \label{fig:Inference Nano GPT}
\end{figure*}

\begin{figure*}[htbp]
  \centering
  \includegraphics[width=\textwidth]{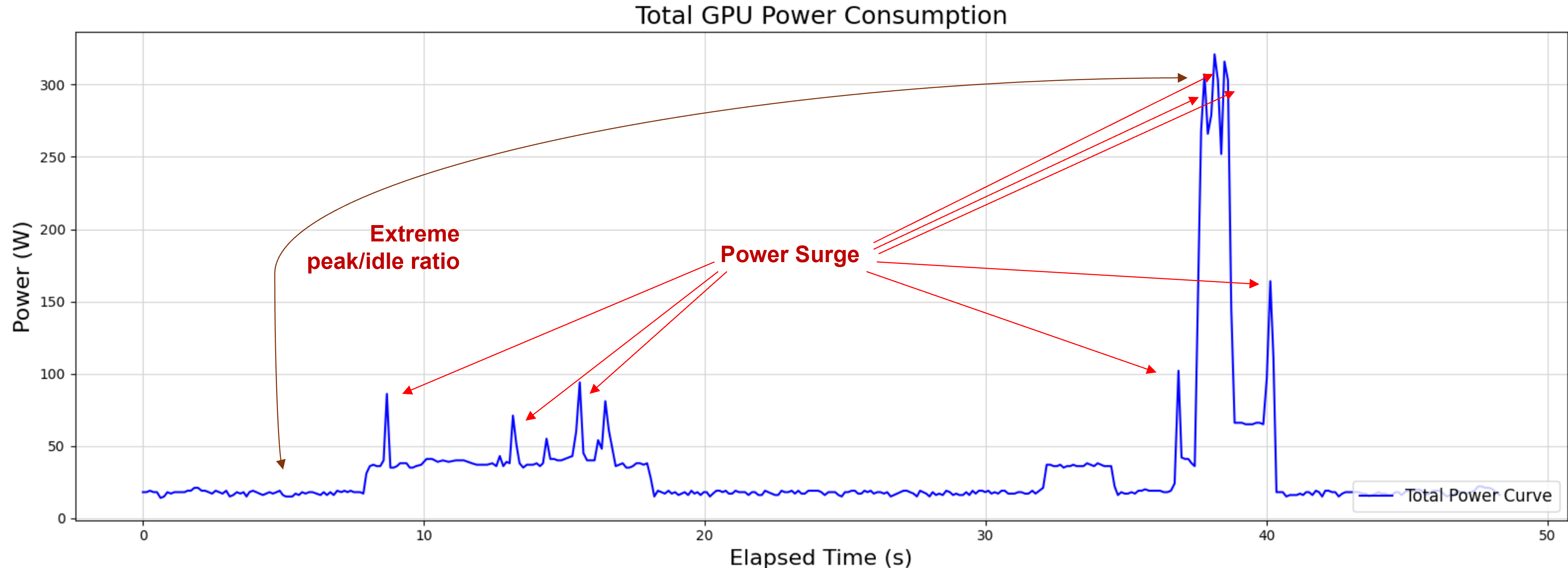}
  \caption{\small Power Consumption of Inference of GPT-2 medium running on setup 2.}
  \label{fig:Inference GPT-2 Medium}
\end{figure*}

The extreme peak-to-idle power ratio observed in both models highlights the dynamic nature of LLM inference tasks as shown in Figure \ref{fig:Inference Nano GPT} and \ref{fig:Inference GPT-2 Medium}. NanoGPT appears to exhibit more frequent but shorter power spikes compared to GPT-2, suggesting differences in their information processing approaches. The maximum energy consumption recorded is 1.57kJ, reflecting the intensive but brief nature of these computational bursts.

These findings have important implications for system design in LLM deployment. The rapid fluctuations between low and high power states pose great challenges for handling swift load changes. Additionally, thermal management systems must be engineered to cope with short but intense heat generation periods.

\subsection{Case study 5: LLM inference with different batch sizes}
The models compared in this power consumption analysis are the state-spaces/mamba-2.8b and EleutherAI/gpt-neo-2.7B \cite{song2024mamba}. The Mamba model, with 2.8 billion parameters, implements the novel selective State Space Model (SSM) architecture designed for efficient sequence processing. The Transformer model, GPT-Neo-2.7B, is based on the GPT architecture and uses the more traditional attention mechanism. Both models have a similar parameter count, allowing for a fair comparison of their respective architectures' impact on power consumption and scalability. This comparison using comparable model sizes highlights the intrinsic differences in how Mamba and Transformer architectures handle increasing computational loads, providing valuable insights into their relative efficiencies and practical limitations.

\begin{figure*}[htbp]
  \centering
  \includegraphics[width=\textwidth]{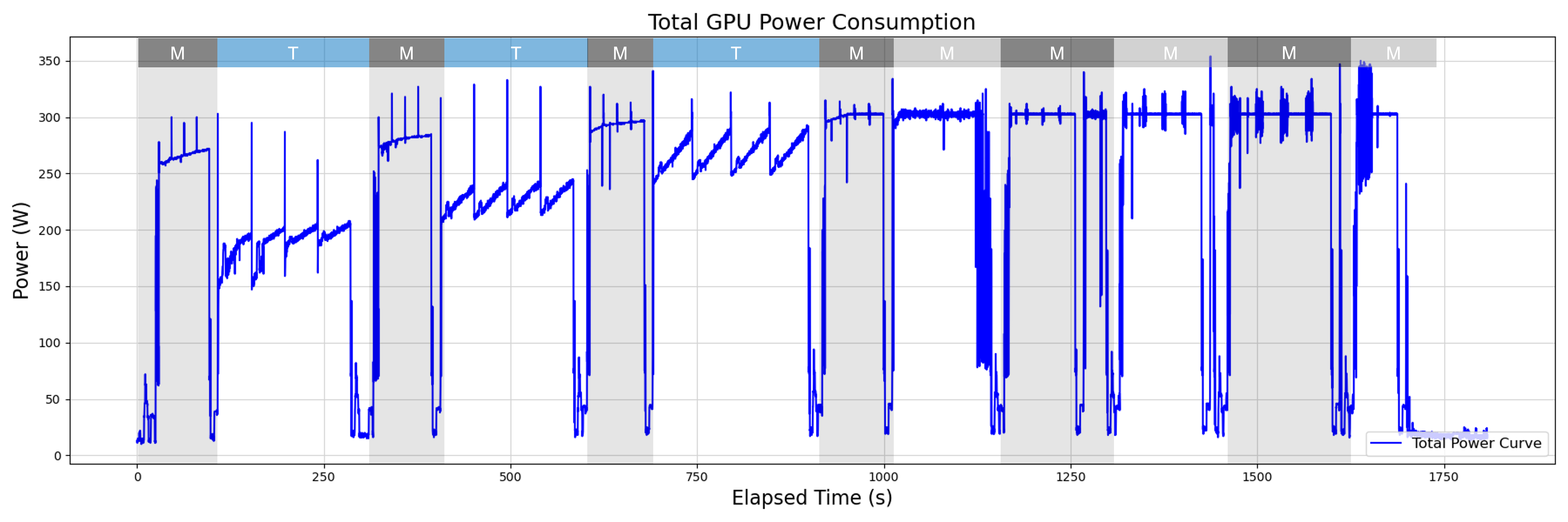}
  \caption{\small Power Consumption of Inference of Mamba-2.8B (grey region) and GPT-Neo-2.7B models (blue region) with the same hardware setup 2. The batch sizes used were 1, 2, 4, 8, 16, 32, 64, and 128 for both models.}
  \label{fig:Inference Mamba-Tranformer}
\end{figure*}

This power consumption graph serves as an illuminating case study on how model architecture and batch size impact GPU power usage during inference as shown in Figure \ref{fig:Inference Mamba-Tranformer}. The alternating gray (Mamba) and light blue (Transformer) sections clearly demonstrate the different power profiles of these architectures as batch sizes increase from left to right. Initially, both models show similar power consumption patterns, but as batch sizes grow, we observe diverging trends. The Transformer model exhibits rapidly escalating power demands, reflected in the steadily rising peaks within its sections, until it abruptly disappears from the graph - due to memory constraints at higher batch sizes (the details of the inference can be found in the Table \ref{table:Mamba-Transformer}). In contrast, Mamba's power consumption increases more gradually, allowing it to continue operating at larger batch sizes where the Transformer model fails. This effectively illustrates Mamba's superior efficiency and scalability. The graph not only highlights immediate performance differences, but also hints at the long-term energy efficiency and cost implications of choosing between these architectures for large-scale deployments.

%% file: Future_and_Opportunities.tex
Based on the findings of this research, promising directions for future research in AI infrastructure power management are identified covering grid side, AI-user side and data-center side.

\subsection{AI-User Perspective}
\subsubsection{Power-aware AI Training, Fine-tuning, and Inference}
A key opportunity lies in developing power-aware AI algorithms that address the unique energy demands of AI's distinct operational phases. For the training phase, which requires sustained high power over extended periods, future research could explore techniques for distributed training that dynamically adjust computation intensity based on available power resources. This might involve load balancing across multiple nodes to reduce peak power demand and allow for more efficient energy use. The fine-tuning phase, with its intermittent high-power usage, presents an opportunity for developing adaptive algorithms that can quickly scale computational intensity. For the inference phase, characterized by highly transient and unpredictable power consumption, future work could focus on methods for adaptive precision, allowing AI models to adjust their computational precision based on input complexity and available energy, potentially smoothing out characteristic power consumption spikes.

\subsubsection{Transparent/Explainable AI Load}
There's significant potential in creating sophisticated tools for visualizing and interpreting power consumption patterns specific to each phase of AI operation. Future developments in this area could produce tools that identify sustained high-power usage periods during training, highlight the intermittent nature of power consumption in fine-tuning, and capture rapid fluctuations in power demand during inference. Another valuable direction would be the development of metrics to quantify energy efficiency across all three AI phases, enabling meaningful comparisons between different AI models and architectures from an energy perspective. Such metrics could drive the field towards more sustainable practices that consider the transient nature of AI power consumption.

\subsubsection{Robust AI Deployment}
Future research could focus on developing robust AI deployment strategies that maintain functionality and performance under unstable power conditions. This is particularly crucial for inference workloads that may experience rapid power fluctuations. Potential areas of exploration include methods for graceful degradation, allowing AI systems to scale back operations under power constraints while maintaining essential services. For training and fine-tuning, future work might examine advanced checkpointing strategies that quickly save and resume progress in response to power fluctuations. The exploration of federated learning approaches also presents an opportunity to distribute power loads across multiple devices or nodes, potentially creating more resilient AI systems.

\subsubsection{Green AI Algorithms}
A promising direction lies in developing AI architectures that inherently require less computational power across all operational phases. Future research could focus on more efficient algorithms for training that achieve comparable results with fewer iterations, thus reducing high-power consumption periods. For fine-tuning, transfer learning techniques that adapt pre-trained models more efficiently could be explored. In the realm of inference, the development of lightweight models for edge devices with limited power resources presents a significant opportunity. Additionally, emerging AI algorithms like KAN \cite{wang2024black}, LNN \cite{Mughees_2024} and more can serve as candidates with less resource requirement. Hardware side, phonic chips represents a promising way to achieve highly efficient AI implementation, and neuromorphic computing approaches mimicking the energy efficiency of biological neural networks could offer dramatic reductions in power consumption, particularly for inference tasks.

\subsection{Data-Center Perspective}
\subsubsection{AI-aware Power Pattern Modeling}
Understanding power consumption patterns across different AI pipeline stages is crucial for designing efficient power delivery and cooling systems in AI-focused data centers. It allows for more precise capacity planning, better load balancing, and the implementation of stage-specific power optimization strategies.

In the pre-training stage, power delivery systems must be capable of sustaining maximum load for extended periods. This requires robust infrastructure capable of handling prolonged high-power draw and the associated thermal management challenges.

The fine-tuning stage may allow for some power optimization, but still requires high performance. Power management strategies in this stage might involve dynamic scaling of resources based on the specific fine-tuning task requirements.

Inference stages offer the most opportunity for power saving through techniques like Dynamic Voltage and Frequency Scaling (DVFS) and selective activation of compute units. The intermittent nature of inference workloads allows for more aggressive power-saving measures during idle periods, while still maintaining low-latency response capabilities for active queries.

By tailoring power management strategies to each stage of the AI pipeline, data centers can significantly improve their energy efficiency while maintaining the high performance required for AI workloads. This stage-specific approach might be formalized as:

\begin{equation}
    P_\text{optimized}(t) = f_\text{stage}(P_\text{total}(t), s(t));
\end{equation}
Where $f_\text{stage}$ is a stage-specific optimization function and $s(t)$ represents the current stage of the AI pipeline at time $t$.

Future developments in this area may focus on more granular modeling of specific AI architectures, integration of renewable energy sources, and advanced power management strategies leveraging AI itself for optimization. These advancements will be crucial as AI workloads continue to grow in scale and complexity, demanding ever more efficient and responsive power management solutions in data center environments.

A potential direction for future research might involve the development of a comprehensive power management function:

\begin{equation}
    P_\text{future}(t) = g(P_\text{total}(t), s(t), R(t), A(t));
\end{equation}
Where $R(t)$ represents the availability of renewable energy sources and $A(t)$ represents the state of an AI-driven power optimization system.

\subsubsection{Hierarchical Operation Management}
\begin{figure}[htbp]
  \centering
  \includegraphics[width=\columnwidth]{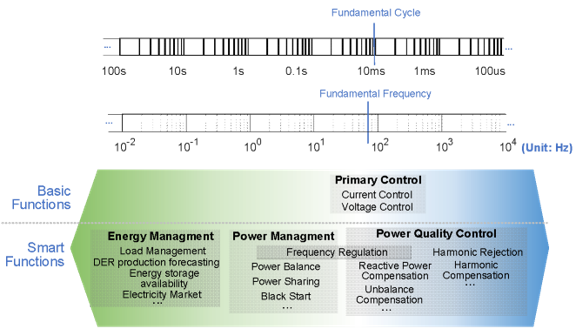}
  \caption{\small The operation management over wide time scale can borrow similar ideas from a smart microgrid.}
  \label{fig:MG}
\end{figure}

Data centers hosting AI workloads face major challenges due to extreme power fluctuations across pre-training, fine-tuning, and inference stages. This variability ranges from sustained high loads during pre-training to rapid, millisecond-level changes in inference, putting significant strain on power delivery systems. To tackle this, data centers in the future can implement a hierarchical operation management approach by following a similar strategy to smart microgrids as summarized in Figure \ref{fig:MG} \cite{li2022smart}. At the macro level, energy management forecasts long-term demands for extended pre-training runs. The mid-tier handles power balancing for variable fine-tuning loads. The fastest control layer deals with power quality, crucial for the split-second spikes of inference tasks. This setup lets operators use techniques like DVFS to maintain stability during rapid load changes. By adopting this multi-tiered strategy, data centers can optimize resource use across all AI stages, fine-tune thermal management, and dynamically adapt to each workload's unique power profile. The result is enhanced performance and energy efficiency, while effectively managing the complex, transient nature of AI power consumption patterns.

\subsubsection{Power Ramping/Decline Compensation}
\begin{figure}[htbp]
  \centering
  \includegraphics[width=\columnwidth]{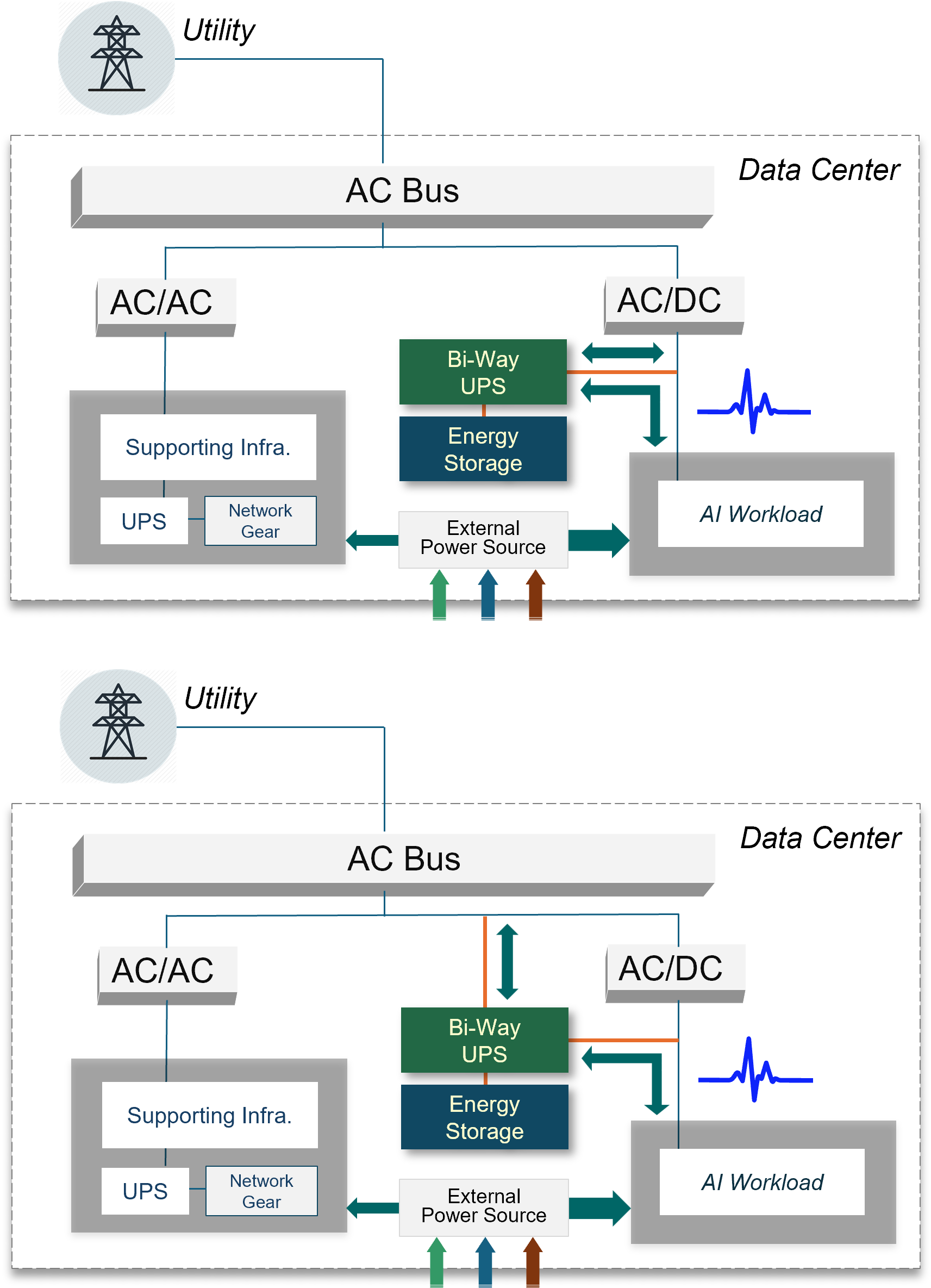}
  \caption{\small Bi-way UPS with capability to smooth the power transience. (Top) UPS is connected to DC bus; (Bottom) UPS is connected to both DC and AC bus.}
  \label{fig:UPS}
\end{figure}

In addition to power management, new devices can also be designed to ease the power surge/dip problems. In Figure \ref{fig:UPS}, we demonstrate a possible solution for the power transience issues in AI-centric data centers. In detail, such a device will have dual functions, which will help handle both the grid side and the data center side. We call it Bi-way uninterruptible power supply (UPS), and it works as a fast-response energy buffer to smooth the power transition from AI power fluctuations to protect the local grid and provide an uninterruptible power supply to the AI facility when the regional grid blackouts. A dedicated energy storage will be installed as traditional UPS but should offer a fast and large enough response. The power compensation function in Bi-way UPS will be triggered if the AI workload ramps too fast or declines too fast. A too-high ramping will stress the local grid and drain the power from the nearby area. The trickier situation is the too-fast decline of the AI load, which may not be predictable and will cause local power hoarding, as the power plant will not be fast enough to adjust the power output and this will lead to a regional power surge.

In Figure \ref{fig:UPS}, two examples are given with slightly different configurations: one is showing with Bi-way UPS connected to the DC bus and works as a parallel compensator for power transience; another one is showing with Bi-way UPS connected with both the AC bus and the DC bus and do the power balance. It should be noted that more configurations are possible and maybe even better than what has been shown here. We treat this issue as an open question and encourage interested readers to explore more solutions.

\subsubsection{Advanced Power Supply and Cooling}
Future power supply and cooling systems could be designed to meet the intense and fast-changing computational requirements of AI systems across their operational phases. For the high-intensity, sustained power draw of training phases, two-phase immersion cooling techniques could be developed to efficiently manage heat from densely packed AI accelerators~\cite{wang2013two}. Adaptive cooling systems that quickly respond to heat generation changes could be created for the variable loads of fine-tuning and inference. This might involve the use of phase-change materials to smooth out thermal fluctuations. On the power supply side, AI-optimized Power Distribution Units (PDUs) with built-in intelligence for dynamic power allocation could be developed. These advanced PDUs could rapidly adjust power distribution based on real-time monitoring of AI workload phases. The integration of wide-bandgap semiconductors in power conversion equipment also presents an opportunity to improve efficiency and reduce heat generation.

\subsection{Power Grid Perspective}
\subsubsection{Regulation of Large-Scale AI}
Future regulatory frameworks could be developed to account for the unique power consumption characteristics of AI systems across training, fine-tuning, and inference phases. These frameworks may include methods for integrating varying power consumption patterns into grid management strategies. For instance, future regulations could require AI operators to provide advance notifications on large-scale training operations or implement power caps during peak demand hours while trading off for critical AI/ML inference tasks. Guidelines for managing unpredictable power fluctuations associated with inference workloads could be established, potentially including requirements for on-site power smoothing technologies. The development of incentive structures to encourage energy-efficient AI practices across all operational phases presents another important opportunity.

\subsubsection{Proactive AI Load Prediction}
Future research could focus on developing sophisticated machine learning models capable of forecasting AI-related power consumption based on the specific characteristics of each operational phase. These models could aim to predict not just overall power demand, but also rapid fluctuations characteristic of AI workloads, particularly during inference. For training phases, models could focus on anticipating the start and duration of sustained high-power usage periods. Fine-tuning predictions might forecast intermittent power demand spikes. Predicting the highly variable power consumption during inference presents a particularly challenging opportunity, requiring models to account for factors like query volume, complexity, and potential batch processing. Integration of these predictive capabilities into real-time monitoring systems for large-scale AI operations could provide grid operators with valuable insights for proactive grid management.

\subsubsection{Renewables Sitting and Sizing}
Future research shall explore optimal strategies for co-locating AI data centers with renewable energy installations, aiming to align power-intensive training phases with periods of peak renewable generation as well as reducing AI-associated carbon emissions. Advanced models could be developed for sizing renewable energy installations based on the expected distribution of AI workloads across training, fine-tuning, and inference phases. For instance, solar installation sizing can be optimized to support daytime inference loads, while wind power could be leveraged for overnight training sessions. The impact of AI power consumption patterns on renewable energy system design and operation presents another area for investigation, potentially leading to the development of hybrid systems that provide stable power despite variability in both AI loads and renewable generation.

\subsubsection{Distribution Grid Planning}
New planning tools could be developed to account for the potentially clustered nature of AI operations and their distinctive power consumption patterns across operational phases. Future research might investigate the impact of concentrated AI workloads on local grid infrastructure and stability, particularly in areas that might see rapid growth in AI-driven industries. Planning tools could consider scenarios where multiple AI systems in a local area simultaneously enter high-power training phases or experience coordinated spikes in inference requests. Another important area for future work is the optimization of power quality in areas with high concentrations of AI workloads, exploring advanced power electronics solutions to mitigate effects of rapid power demand fluctuations.

\subsubsection{Demand-Side Management}
Future demand response programs could be specifically tailored to the different phases of AI operations. For training workloads, strategies might involve scheduling these power-intensive tasks during periods of excess renewable energy generation or lower overall grid demand. Fine-tuning operations could be managed through dynamic pricing schemes that incentivize spreading these tasks out to avoid coincident demand spikes. For inference workloads, real-time demand response mechanisms could be developed to quickly adjust AI computation intensity based on momentary grid conditions. This might involve creating APIs that allow grid operators to send rapid signals to AI systems, requesting brief reductions in power consumption during critical periods.

\subsubsection{Grid-Scale Energy Storage}
Future grid-scale energy storage solutions could be designed to address the variability and intensity of AI workloads across training, fine-tuning, and inference phases. Hybrid energy storage systems combining multiple technologies could be developed to meet the diverse needs of AI power consumption patterns. Long-term storage solutions might be explored for managing the sustained, high-power demands of training phases, while fast-response technologies could be integrated to handle fine-tuning spikes and inference fluctuations. Control strategies for these storage systems could be designed to anticipate and respond to the characteristic power profiles of different AI operational phases.

\subsubsection{Fast Frequency Control}
To address the challenges posed by rapid power fluctuations in AI workloads, future research could focus on developing ultra-fast response frequency regulation techniques. This might include exploring the use of AI accelerators themselves as frequency regulation resources, leveraging their ability to rapidly adjust power consumption. The potential for using AI computation as a form of synthetic inertia could be investigated, where AI systems might be programmed to respond to rate-of-change-of-frequency events, mimicking the inertial response of traditional generators.

\subsubsection{AI for Grid Stabilization}
Future research could explore how AI infrastructure, particularly during less intensive inference periods or idle times between training sessions, might be leveraged for grid stabilization services. This could involve using AI systems for voltage support and reactive power compensation. The development of AI-driven grid optimization techniques for real-time power flow management presents another promising direction. Future work might also investigate the potential for creating virtual power plants composed of distributed AI computing resources, which could provide significant demand response capabilities and help balance supply and demand on the grid at various timescales.

\subsubsection{Consumer Adoption and Education}
As AI technologies become more prevalent in consumer devices, future work could focus on developing energy efficiency ratings and labeling systems specifically for AI-enabled consumer devices. These systems might take into account energy consumption when adopting AI, helping consumers make informed decisions about the energy implications of their AI-powered products. Educational programs could be developed to inform the public about the energy implications of AI technologies and best practices for their sustainable use, including guidance on energy-efficient settings for AI-powered devices and the impact of always-on inference capabilities on energy consumption.

%% file: Appendix.tex
\subsection{Analysis of GPU Power Consumption During Fine-Tuning}
\label{subsec:Appendix-Inference}

\textbf{Background:}
In a detailed guide from AMD \cite{song2024mamba}, the process of implementing and training a Generative Pre-trained Transformer (GPT) model using JAX on AMD GPUs is elucidated. The guide draws inspiration from Andrej Karpathy's PyTorch-based nanoGPT and discusses converting these models for use with JAX/Flax. The tutorial leverages the character-level Shakespeare dataset to illustrate the training of these lightweight GPT models. Here we breakdown the details about the unique power consumption behaviors during different stages of the fine-tuning process of the GPT-medium based on this project.

\subsubsection{Time Window: 0-350s}
\textbf{Stage: Initialization and Setup}

\textbf{Relevant Code Sections:}
\begin{lstlisting}
if init_from == 'scratch':
    # Initialize a new model from scratch
    gptconf = GPTConfig(**model_args)
    model = GPT(gptconf)
    params = model.init(
        jax.random.PRNGKey(1), idx)
elif init_from == "resume":
    # Resuming training from a checkpoint
    ...
X, Y = get_batch('train') 
# fetch the very first batch
state, lr_schedule = init_train_state(
    model, params['params'], learning_rate, 
    weight_decay, beta1, beta2, decay_lr, 
    warmup_iters, lr_decay_iters, min_lr)
\end{lstlisting}

\textbf{What Happens:} The model initialization occurs either from scratch or from a saved checkpoint, involving the configuration and instantiation of the model along with initial parameter setting. Concurrently, the first data batch is fetched, setting the stage for the training process. The optimizer and learning rate schedules are initialized, which are critical for controlling the training dynamics.

\textbf{Power Consumption:} The initiation of these processes results in an initial surge in power consumption, indicative of the GPU's increasing operational load. This rise is particularly noticeable as memory allocations take place, pre-trained weights are loaded if applicable, and initial data processing begins.

\subsubsection{Time Window: 350-1700s}
\textbf{Stage: Early Training and Learning Rate Warmup}

\textbf{Relevant Code Sections:}
\begin{lstlisting}
lr_schedule = optax.warmup_cosine_decay_schedule(
    init_value=1e-9, peak_value=learning_rate,
    warmup_steps=warmup_iters, 
    decay_steps=lr_decay_iters, end_value=min_lr)

while True:
    if iter_num % eval_interval == 0:
        print(f'Evaluating at iter_num == {iter_num}...')
        losses = estimate_loss(state)
        # (Save checkpoints, log losses)
    state, loss, rng_key = train_step(
        state, get_batch('train'), rng_key)
\end{lstlisting}

\textbf{What Happens:} This phase marks the beginning of active training. The learning rate is in its warmup period, gradually increasing to its peak value. Each iteration involves forward and backward passes for loss computation and parameter updates. Periodic evaluations and checkpoints occur as defined by \texttt{eval\_interval}.

\textbf{Power Consumption:} This stage is characterized by high and fluctuating power draw, driven by the intensive computational demands of training. The increasing learning rate during the warmup period may lead to more aggressive parameter updates, potentially resulting in more erratic fluctuations in power consumption. Periodic drops in power usage are observed, likely corresponding to evaluation intervals.

\subsubsection{Time Window: 1700-3100s}
\textbf{Stage: Late Training and Learning Rate Decay}

\textbf{Relevant Code Sections:}
\begin{lstlisting}
lr_schedule = optax.warmup_cosine_decay_schedule(
    init_value=1e-9, peak_value=learning_rate,
    warmup_steps=warmup_iters, 
    decay_steps=lr_decay_iters, end_value=min_lr)

state, loss, rng_key = train_step(
    state, get_batch('train'), rng_key)
if iter_num % log_interval == 0:
    print(f"iter {iter_num}: loss {loss:.4f}, time {dt*1000:.2f}ms")
\end{lstlisting}

\textbf{What Happens:} The training continues with consistent iterations, but now the learning rate has entered its decay phase. This phase involves more refined parameter updates as the learning rate gradually decreases. Regular logging of metrics helps in monitoring the model's performance closely.

\textbf{Power Consumption:} The power usage during this phase remains high and fluctuating, similar to the previous stage. However, there might be subtle differences in the pattern of fluctuations. The decaying learning rate may lead to slightly more consistent power consumption patterns, although this difference may not be dramatically visible in the power graph. Periodic drops in power consumption continue, corresponding to evaluation intervals.

\subsubsection{Time Window: 3100-3200s}
\textbf{Stage: Training Completion and Shutdown}

\textbf{Relevant Code Sections:}
\begin{lstlisting}
if iter_num > max_iters:
    break
\end{lstlisting}

\textbf{What Happens:} As the training reaches the maximum number of iterations, the process begins to wind down. Final evaluations and checkpointing may occur during this phase.

\textbf{Power Consumption:} A sharp drop in power consumption is observed towards the end of this phase, reflecting the completion of training and the winding down of active GPU engagement.

\subsubsection{Summary}
\begin{itemize}
    \item \textbf{0-350s:} Initialization tasks with an initial spike in power consumption.
    \item \textbf{350-1700s:} Early training with learning rate warmup, characterized by high and potentially more erratic power fluctuations.
    \item \textbf{1700-3100s:} Late training with learning rate decay, showing high but potentially more consistent power fluctuations.
    \item \textbf{3100-3200s:} Training completion with a sharp drop in power consumption.
\end{itemize}

Throughout the main training phase (350-3100s), periodic drops in power consumption are observed, likely corresponding to evaluation intervals as defined by \texttt{eval\_interval} in the code. The learning rate schedule plays a crucial role in defining the training stages, even though its impact on power consumption is subtle and not dramatically visible in the power graph.

\subsection{Different LLMs} 
\label{subsec:Appendix-SLM}

The detailed parameters of the LLMs evaluated in this work are shown in Table \ref{tab:model-specs}.

\begin{table*}[htbp]
\caption{Specifications of Various AI Model Series: Small Language Model \cite{huggingface2024gpt2, karpathy2024nanogpt, statespaces2024mamba, eleutherai2024gptneo}}
\centering
\resizebox{\textwidth}{!}{%
\begin{tabular}{|l|l|r|r|r|r|r|}
\hline
\textbf{Series} & \textbf{Model Name} & \textbf{Parameters} & \textbf{Layers} & \textbf{Hidden Size} & \textbf{Attention Heads} & \textbf{Context Length} \\
\hline
\multirow{4}{*}{GPT-2} 
 & GPT-2 Small & 124M & 12 & 768 & 12 & 1024 \\
 & GPT-2 Medium & 355M & 24 & 1024 & 16 & 1024 \\
 & GPT-2 Large & 774M & 36 & 1280 & 20 & 1024 \\
 & GPT-2 XL & 1.5B & 48 & 1600 & 25 & 1024 \\
\hline
\multirow{2}{*}{nanoGPT} 
 & nanoGPT (base) & Varies & Varies & Varies & Varies & Configurable \\
 & nanoGPT-124M & 124M & 12 & 768 & 12 & Configurable \\
\hline
\multirow{5}{*}{Mamba} 
 & mamba-130M & 130M & 24 & 768 & N/A & 2048 \\
 & mamba-370M & 370M & 48 & 1024 & N/A & 2048 \\
 & mamba-790M & 790M & 48 & 1536 & N/A & 2048 \\
 & mamba-1.4B & 1.4B & 48 & 2048 & N/A & 2048 \\
 & mamba-2.8B & 2.8B & 64 & 2560 & N/A & 2048 \\
\hline
\multirow{4}{*}{GPT-Neo} 
 & GPT-Neo 125M & 125M & 12 & 768 & 12 & 2048 \\
 & GPT-Neo 1.3B & 1.3B & 24 & 2048 & 16 & 2048 \\
 & GPT-Neo 2.7B & 2.7B & 32 & 2560 & 20 & 2048 \\
 & GPT-NeoX 20B & 20B & 44 & 6144 & 64 & 2048 \\
\hline
\end{tabular}
}
\label{tab:model-specs}
\end{table*}

\subsection{GPU Specification}

The detailed specifications of the GPUs that relate to case studies are listed in Table \ref{tab:gpu-comparison}.

\begin{table*}[h]
\centering
\caption{Comparison of High-Performance GPUs \cite{nvidia2024v100, nvidia2024a40, technicalcity2024comparison}}
\label{tab:gpu-comparison}
\resizebox{\textwidth}{!}{%
\begin{tabular}{|l|c|c|c|c|}
\hline
\textbf{Specification} & \textbf{NVIDIA Volta V100} & \textbf{NVIDIA A40} & \textbf{NVIDIA GeForce RTX 4090} & \textbf{AMD Radeon RX 7900 XTX} \\
\hline
Architecture & NVIDIA Volta & NVIDIA Ampere & Ada Lovelace (2022-2024) & RDNA 3 (2022-2023) \\
\hline
CUDA Cores / Stream Processors & 5,120 & 10,752 & 16,384 & 6,144 \\
\hline
Tensor Cores & 640 & 336 (3rd Gen) & 512 (4th Gen) & N/A \\
\hline
Memory & 32 GB / 16 GB HBM2 & 48 GB GDDR6 with ECC & 24 GB GDDR6X & 24 GB GDDR6 \\
\hline
Memory Bandwidth & \begin{tabular}[c]{@{}c@{}}900 GB/sec (PCIe, SXM2),\\ 1134 GB/sec (V100S)\end{tabular} & 696 GB/s & 1,008 GB/s & 960.0 GB/s \\
\hline
Memory Interface & 4096-bit & 384-bit & 384-bit & 384-bit \\
\hline
Boost Clock & 1530 MHz & Not specified & 2,520 MHz & 2,270 MHz \\
\hline
System Interface & \begin{tabular}[c]{@{}c@{}}PCIe Gen3 (PCIe, V100S),\\ NVIDIA NVLink (SXM2)\end{tabular} & PCIe Gen4: 64GB/s & PCIe 4.0 x16 & PCIe 4.0 x16 \\
\hline
Form Factor & \begin{tabular}[c]{@{}c@{}}PCIe Full Height/Length\\ (PCIe, V100S), SXM2\end{tabular} & 4.4" (H) x 10.5" (L) dual slot & Not specified & Not specified \\
\hline
Max Power Consumption & \begin{tabular}[c]{@{}c@{}}250 W (PCIe, V100S),\\ 300 W (SXM2)\end{tabular} & 300 W & 450 W & 355 W \\
\hline
Compute APIs & \begin{tabular}[c]{@{}c@{}}CUDA, DirectCompute,\\ OpenCL, OpenACC\end{tabular} & \begin{tabular}[c]{@{}c@{}}CUDA, DirectCompute,\\ OpenCL, OpenACC\end{tabular} & \begin{tabular}[c]{@{}c@{}}CUDA, DirectCompute,\\ OpenCL, OpenGL\end{tabular} & \begin{tabular}[c]{@{}c@{}}DirectCompute,\\ OpenCL, OpenGL\end{tabular} \\
\hline
Launch Price (MSRP) & \$10,000 & \$5,000 & \$1,599 & \$999 \\
\hline
Release Date & 2017 & October 2020 & September 20, 2022 & November 3, 2022 \\
\hline
Manufacturing Process & 12 nm & 7 nm & 5 nm & 5 nm \\
\hline
Floating-point Performance (FP32) & 15.7 TFLOPS & 37.4 TFLOPS & 82.58 TFLOPS & 61.42 TFLOPS \\
\hline
Memory Clock & 877 MHz (1.75 Gbps effective) & 14,500 MHz & 21,000 MHz & 20,000 MHz \\
\hline
\end{tabular}%
}
\end{table*}

\subsection{Inference Test Details of Mamba and Transformer}

The Mamba model demonstrates excellent scalability with increasing batch sizes, maintaining reasonable inference times and memory usage up to batch sizes as large as 64 before encountering OOM issues at batch size 128. The selective state-space modeling in Mamba allows it to efficiently handle larger sequences with lower memory requirements, which is evident from the more stable inference times and delayed OOM errors compared to the Transformer model. The Transformer model struggles with increasing batch sizes, leading to significant slowdowns and early OOM errors. The memory-intensive nature of attention mechanisms, especially with larger batch sizes, severely impacts its performance on the given hardware (AMD GPU with ROCm).

\begin{table*}[h!]
\caption{Comparison of Mamba and GPT-Neo models across different batch sizes. OOM indicates an Out Of Memory error.}
\centering
\begin{tabular}{|l|c|c|c|c|c|}
\hline
\textbf{Model Name} & \textbf{Parameters} & \textbf{Batch Size} & \textbf{Prompt Length} & \textbf{Processing Time (s)} & \textbf{GPU Memory Used (GB)} \\ \hline

\multirow{8}{*}{\textbf{Mamba-2.8B}} & \multirow{8}{*}{2,768,345,600} & 1   & \multirow{8}{*}{1024} & 16.86 & 5  \\ \cline{3-3} \cline{5-6}
                                      &                                 & 2   &                        & 17.60 & 6  \\ \cline{3-3} \cline{5-6}
                                      &                                 & 4   &                        & 18.23 & 6  \\ \cline{3-3} \cline{5-6}
                                      &                                 & 8   &                        & 19.50 & 7  \\ \cline{3-3} \cline{5-6}
                                      &                                 & 16  &                        & 22.07 & 9  \\ \cline{3-3} \cline{5-6}
                                      &                                 & 32  &                        & 26.27 & 12 \\ \cline{3-3} \cline{5-6}
                                      &                                 & 64  &                        & 33.51 & 18 \\ \cline{3-3} \cline{5-6}
                                      &                                 & 128 &                        & OOM   & OOM \\ \hline

\multirow{8}{*}{\textbf{GPT-Neo-2.7B (Eager)}} & \multirow{8}{*}{2,651,307,520} & 1   & \multirow{8}{*}{1024} & 44.15 & 7  \\ \cline{3-3} \cline{5-6}
                                               &                                 & 2   &                        & 44.43 & 9  \\ \cline{3-3} \cline{5-6}
                                               &                                 & 4   &                        & 52.07 & 13 \\ \cline{3-3} \cline{5-6}
                                               &                                 & 8   &                        & OOM   & OOM \\ \cline{3-3} \cline{5-6}
                                               &                                 & 16  &                        & OOM   & OOM \\ \cline{3-3} \cline{5-6}
                                               &                                 & 32  &                        & OOM   & OOM \\ \cline{3-3} \cline{5-6}
                                               &                                 & 64  &                        & OOM   & OOM \\ \cline{3-3} \cline{5-6}
                                               &                                 & 128 &                        & OOM   & OOM \\ \hline

\end{tabular}
\label{table:Mamba-Transformer}
\end{table*}

%% file: main_arxiv.bbl
\begin{thebibliography}{10}
\providecommand{\url}[1]{#1}
\csname url@samestyle\endcsname
\providecommand{\newblock}{\relax}
\providecommand{\bibinfo}[2]{#2}
\providecommand{\BIBentrySTDinterwordspacing}{\spaceskip=0pt\relax}
\providecommand{\BIBentryALTinterwordstretchfactor}{4}
\providecommand{\BIBentryALTinterwordspacing}{\spaceskip=\fontdimen2\font plus
\BIBentryALTinterwordstretchfactor\fontdimen3\font minus \fontdimen4\font\relax}
\providecommand{\BIBforeignlanguage}[2]{{%
\expandafter\ifx\csname l@#1\endcsname\relax
\typeout{** WARNING: IEEEtran.bst: No hyphenation pattern has been}%
\typeout{** loaded for the language `#1'. Using the pattern for}%
\typeout{** the default language instead.}%
\else
\language=\csname l@#1\endcsname
\fi
#2}}
\providecommand{\BIBdecl}{\relax}
\BIBdecl

\bibitem{bommasani2021opportunities}
R.~Bommasani, D.~A. Hudson, E.~Adeli, R.~Altman, S.~Arora, S.~von Arx, M.~S. Bernstein, J.~Bohg, A.~Bosselut, E.~Brunskill \emph{et~al.}, ``On the opportunities and risks of foundation models,'' \emph{arXiv preprint arXiv:2108.07258}, 2021.

\bibitem{bubeck2023sparks}
S.~Bubeck, V.~Chandrasekaran, R.~Eldan, J.~Gehrke, E.~Horvitz, E.~Kamar, P.~Lee, Y.~T. Lee, Y.~Li, S.~Lundberg \emph{et~al.}, ``Sparks of artificial general intelligence: Early experiments with gpt-4,'' \emph{arXiv preprint arXiv:2303.12712}, 2023.

\bibitem{dubey2024llama}
A.~Dubey, A.~Jauhri, A.~Pandey, A.~Kadian, A.~Al-Dahle, A.~Letman, A.~Mathur, A.~Schelten, A.~Yang, A.~Fan \emph{et~al.}, ``The llama 3 herd of models,'' \emph{arXiv preprint arXiv:2407.21783}, 2024.

\bibitem{iea2024electricity}
{International Energy Agency}, ``Electricity 2024,'' \url{https://www.iea.org/reports/electricity-2024}, 2024, iEA, Paris. Licence: CC BY 4.0. Accessed: 2024-09-07.

\bibitem{touvron2023llama}
H.~Touvron, L.~Martin, K.~Stone, P.~Albert, A.~Almahairi, Y.~Babaei, N.~Bashlykov, S.~Batra, P.~Bhargava, S.~Bhosale \emph{et~al.}, ``Llama 2: Open foundation and fine-tuned chat models,'' \emph{arXiv preprint arXiv:2307.09288}, 2023.

\bibitem{achiam2023gpt}
J.~Achiam, S.~Adler, S.~Agarwal, L.~Ahmad, I.~Akkaya, F.~L. Aleman, D.~Almeida, J.~Altenschmidt, S.~Altman, S.~Anadkat \emph{et~al.}, ``Gpt-4 technical report,'' \emph{arXiv preprint arXiv:2303.08774}, 2023.

\bibitem{lui2021understanding}
M.~Lui, Y.~Yetim, {\"O}.~{\"O}zkan, Z.~Zhao, S.-Y. Tsai, C.-J. Wu, and M.~Hempstead, ``Understanding capacity-driven scale-out neural recommendation inference,'' in \emph{2021 IEEE International Symposium on Performance Analysis of Systems and Software (ISPASS)}.\hskip 1em plus 0.5em minus 0.4em\relax IEEE, 2021, pp. 162--171.

\bibitem{kaack2022aligning}
L.~H. Kaack, P.~L. Donti, E.~Strubell, G.~Kamiya, F.~Creutzig, and D.~Rolnick, ``Aligning artificial intelligence with climate change mitigation,'' \emph{Nature Climate Change}, vol.~12, no.~6, pp. 518--527, 2022.

\bibitem{forbes2024gpt}
Forbes, ``Ai is pushing the world toward an energy crisis,'' \url{https://www.forbes.com/sites/arielcohen/2024/05/23/ai-is-pushing-the-world-towards-an-energy-crisis/}, 2024, accessed: 2024-09-07.

\bibitem{kirby2005method}
B.~Kirby and M.~Milligan, ``Method and case study for estimating the ramping capability of a control area or balancing authority and implications for moderate or high wind penetration,'' National Renewable Energy Lab., Golden, CO (US), Tech. Rep., 2005.

\bibitem{sevilla2022compute}
J.~Sevilla, L.~Heim, A.~Ho, T.~Besiroglu, M.~Hobbhahn, and P.~Villalobos, ``Compute trends across three eras of machine learning,'' in \emph{2022 International Joint Conference on Neural Networks (IJCNN)}.\hskip 1em plus 0.5em minus 0.4em\relax IEEE, 2022, pp. 1--8.

\bibitem{masanet2024better}
E.~Masanet, N.~Lei, and J.~Koomey, ``To better understand ai’s growing energy use, analysts need a data revolution,'' \emph{Joule}, 2024.

\bibitem{crawford2024generative}
\BIBentryALTinterwordspacing
K.~Crawford, ``Generative ai’s environmental costs are soaring — and mostly secret,'' \emph{Nature}, vol. 626, p. 693, February 2024, wORLD VIEW, First-of-its-kind US bill would address the environmental costs of the technology, but there’s a long way to go. [Online]. Available: \url{https://doi.org/10.1038/d41586-024-00478-x}
\BIBentrySTDinterwordspacing

\bibitem{de2023growing}
A.~de~Vries, ``The growing energy footprint of artificial intelligence,'' \emph{Joule}, vol.~7, no.~10, pp. 2191--2194, 2023.

\bibitem{avelar2023ai}
V.~Avelar, P.~Donovan, P.~Lin, W.~Torell, and M.~A.~T. Arango, ``The ai disruption: Challenges and guidance for data center design,'' \emph{Schneider Electric [Online]}, 2023.

\bibitem{guo2024bridge}
J.~Guo, A.~Pickthall, T.-L. Shih, and L.~Floridi, ``How to bridge the us energy supply gap to meet the rising demands for computing power in the era of generative ai?'' Available at SSRN: \url{https://ssrn.com/abstract=4887664} or \url{http://dx.doi.org/10.2139/ssrn.4887664}, May 2024, accessed: 2024-09-07.

\bibitem{li2022ai}
B.~Li, R.~Arora, S.~Samsi, T.~Patel, W.~Arcand, D.~Bestor, C.~Byun, R.~B. Roy, B.~Bergeron, J.~Holodnak \emph{et~al.}, ``Ai-enabling workloads on large-scale gpu-accelerated system: Characterization, opportunities, and implications,'' in \emph{2022 IEEE International Symposium on High-Performance Computer Architecture (HPCA)}.\hskip 1em plus 0.5em minus 0.4em\relax IEEE, 2022, pp. 1224--1237.

\bibitem{bianchini2024datacenter}
R.~Bianchini, C.~Belady, and A.~Sivasubramaniam, ``Datacenter power and energy management: past, present, and future,'' \emph{IEEE Micro}, 2024.

\bibitem{zhang2018load}
J.~Zhang, F.~R. Yu, S.~Wang, T.~Huang, Z.~Liu, and Y.~Liu, ``Load balancing in data center networks: A survey,'' \emph{IEEE Communications Surveys \& Tutorials}, vol.~20, no.~3, pp. 2324--2352, 2018.

\bibitem{radovanovic2021power}
A.~Radovanovic, B.~Chen, S.~Talukdar, B.~Roy, A.~Duarte, and M.~Shahbazi, ``Power modeling for effective datacenter planning and compute management,'' \emph{IEEE Transactions on Smart Grid}, vol.~13, no.~2, pp. 1611--1621, 2021.

\bibitem{he2024long}
X.~He, D.~H. Tsang, and Y.~Chen, ``Long-term carbon-efficient planning for geographically shiftable resources: A monte carlo tree search approach,'' \emph{IEEE Transactions on Power Systems}, 2024.

\bibitem{luccioni2023estimating}
A.~S. Luccioni, S.~Viguier, and A.-L. Ligozat, ``Estimating the carbon footprint of bloom, a 176b parameter language model,'' \emph{Journal of Machine Learning Research}, vol.~24, no. 253, pp. 1--15, 2023.

\bibitem{mlenergy2024leaderboard}
{ML Energy}, ``Ml energy leaderboard,'' \url{https://ml.energy/leaderboard/?__theme=light}, 2024, accessed: 2024-09-07.

\bibitem{llama2024herd}
\BIBentryALTinterwordspacing
A.~Llama~Team, ``The llama 3 herd of models,'' July 2024, a detailed contributor list can be found in the appendix of this paper. [Online]. Available: \url{https://llama.meta.com/}
\BIBentrySTDinterwordspacing

\bibitem{elmeleegy2024demystifying}
A.~Elmeleegy, S.~Raj, B.~Slechta, and V.~Mehta, ``Demystifying ai inference deployments for trillion parameter large language models,'' \url{https://developer.nvidia.com/blog/demystifying-ai-inference-deployments-for-trillion-parameter-large-language-models/}, June 2024, technical Blog, Data Center / Cloud, NVIDIA. Accessed: 2024-09-07.

\bibitem{mcdonald2022great}
J.~McDonald, B.~Li, N.~Frey, D.~Tiwari, V.~Gadepally, and S.~Samsi, ``Great power, great responsibility: Recommendations for reducing energy for training language models,'' \emph{arXiv preprint arXiv:2205.09646}, 2022.

\bibitem{10363447}
S.~Samsi, D.~Zhao, J.~McDonald, B.~Li, A.~Michaleas, M.~Jones, W.~Bergeron, J.~Kepner, D.~Tiwari, and V.~Gadepally, ``From words to watts: Benchmarking the energy costs of large language model inference,'' in \emph{2023 IEEE High Performance Extreme Computing Conference (HPEC)}, 2023, pp. 1--9.

\bibitem{patel2024characterizing}
P.~Patel, E.~Choukse, C.~Zhang, {\'I}.~Goiri, B.~Warrier, N.~Mahalingam, and R.~Bianchini, ``Characterizing power management opportunities for llms in the cloud,'' in \emph{Proceedings of the 29th ACM International Conference on Architectural Support for Programming Languages and Operating Systems, Volume 3}, 2024, pp. 207--222.

\bibitem{hu2024characterization}
Q.~Hu, Z.~Ye, Z.~Wang, G.~Wang, M.~Zhang, Q.~Chen, P.~Sun, D.~Lin, X.~Wang, Y.~Luo \emph{et~al.}, ``Characterization of large language model development in the datacenter,'' in \emph{21st USENIX Symposium on Networked Systems Design and Implementation (NSDI 24)}, 2024, pp. 709--729.

\bibitem{karimi2024profiling}
A.~M. Karimi, N.~S. Sattar, W.~Shin, and F.~Wang, ``Profiling and modeling of power characteristics of leadership-scale hpc system workloads,'' \emph{arXiv preprint arXiv:2402.00729}, 2024.

\bibitem{ORO2015429}
\BIBentryALTinterwordspacing
E.~Oró, V.~Depoorter, A.~Garcia, and J.~Salom, ``Energy efficiency and renewable energy integration in data centres. strategies and modelling review,'' \emph{Renewable and Sustainable Energy Reviews}, vol.~42, pp. 429--445, 2015. [Online]. Available: \url{https://www.sciencedirect.com/science/article/pii/S1364032114008600}
\BIBentrySTDinterwordspacing

\bibitem{KWON2020115424}
\BIBentryALTinterwordspacing
S.~Kwon, ``Ensuring renewable energy utilization with quality of service guarantee for energy-efficient data center operations,'' \emph{Applied Energy}, vol. 276, p. 115424, 2020. [Online]. Available: \url{https://www.sciencedirect.com/science/article/pii/S0306261920309363}
\BIBentrySTDinterwordspacing

\bibitem{kong2014survey}
F.~Kong and X.~Liu, ``A survey on green-energy-aware power management for datacenters,'' \emph{ACM Computing Surveys (CSUR)}, vol.~47, no.~2, pp. 1--38, 2014.

\bibitem{anthony2020carbontracker}
L.~F.~W. Anthony, B.~Kanding, and R.~Selvan, ``Carbontracker: Tracking and predicting the carbon footprint of training deep learning models,'' \emph{arXiv preprint arXiv:2007.03051}, 2020.

\bibitem{alavani2023program}
G.~Alavani, J.~Desai, S.~Saha, and S.~Sarkar, ``Program analysis and machine learning--based approach to predict power consumption of cuda kernel,'' \emph{ACM Transactions on Modeling and Performance Evaluation of Computing Systems}, vol.~8, no.~4, pp. 1--24, 2023.

\bibitem{cao2023gpu}
J.~Cao, R.~Sen, M.~Interlandi, J.~Arulraj, and H.~Kim, ``Gpu database systems characterization and optimization,'' \emph{Proceedings of the VLDB Endowment}, vol.~17, no.~3, pp. 441--454, 2023.

\bibitem{wu2022sustainable}
C.-J. Wu, R.~Raghavendra, U.~Gupta, B.~Acun, N.~Ardalani, K.~Maeng, G.~Chang, F.~Aga, J.~Huang, C.~Bai \emph{et~al.}, ``Sustainable ai: Environmental implications, challenges and opportunities,'' \emph{Proceedings of Machine Learning and Systems}, vol.~4, pp. 795--813, 2022.

\bibitem{lacoste2019quantifying}
A.~Lacoste, A.~Luccioni, V.~Schmidt, and T.~Dandres, ``Quantifying the carbon emissions of machine learning,'' \emph{arXiv preprint arXiv:1910.09700}, 2019.

\bibitem{henderson2020towards}
P.~Henderson, J.~Hu, J.~Romoff, E.~Brunskill, D.~Jurafsky, and J.~Pineau, ``Towards the systematic reporting of the energy and carbon footprints of machine learning,'' \emph{Journal of Machine Learning Research}, vol.~21, no. 248, pp. 1--43, 2020.

\bibitem{stojkovic2024dynamollm}
J.~Stojkovic, C.~Zhang, {\'I}.~Goiri, J.~Torrellas, and E.~Choukse, ``Dynamollm: Designing llm inference clusters for performance and energy efficiency,'' \emph{arXiv preprint arXiv:2408.00741}, 2024.

\bibitem{devlin2018bert}
J.~Devlin, ``Bert: Pre-training of deep bidirectional transformers for language understanding,'' \emph{arXiv preprint arXiv:1810.04805}, 2018.

\bibitem{jia2019beyond}
Z.~Jia, M.~Zaharia, and A.~Aiken, ``Beyond data and model parallelism for deep neural networks.'' \emph{Proceedings of Machine Learning and Systems}, vol.~1, pp. 1--13, 2019.

\bibitem{google2024datacenters}
Google, ``Data center efficiency,'' \url{https://www.google.com/about/datacenters/efficiency}, 2024, accessed: 2024-09-07.

\bibitem{mcqueen2020department}
S.~McQueen, J.~Stanford, S.~Satyapal, E.~Miller, N.~Stetson, D.~Papageorgopoulos, N.~Rustagi, V.~Arjona, J.~Adams, K.~Randolph \emph{et~al.}, ``Department of energy hydrogen program plan,'' US Department of Energy (USDOE), Washington DC (United States), Tech. Rep., 2020.

\bibitem{salomonsson2007adaptive}
D.~Salomonsson, L.~Soder, and A.~Sannino, ``An adaptive control system for a dc microgrid for data centers,'' in \emph{2007 IEEE Industry Applications Annual Meeting}.\hskip 1em plus 0.5em minus 0.4em\relax IEEE, 2007, pp. 2414--2421.

\bibitem{samsi2021supercloud}
S.~Samsi, M.~L. Weiss, D.~Bestor, B.~Li, M.~Jones, A.~Reuther, D.~Edelman, W.~Arcand, C.~Byun, J.~Holodnack \emph{et~al.}, ``The mit supercloud dataset,'' in \emph{2021 IEEE High Performance Extreme Computing Conference (HPEC)}.\hskip 1em plus 0.5em minus 0.4em\relax IEEE, 2021, pp. 1--8.

\bibitem{huggingface2024gpt2}
H.~Face, ``Gpt-2 model documentation,'' \url{https://huggingface.co/docs/transformers/model_doc/gpt2}, 2024, accessed: 2024-09-07.

\bibitem{karpathy2024nanogpt}
A.~Karpathy, ``nanogpt: A tiny gpt implementation in pytorch,'' \url{https://github.com/karpathy/nanoGPT}, 2024, accessed: 2024-09-07.

\bibitem{jia2024guide}
D.~Jia, ``A guide to implementing and training generative pre-trained transformers (gpt) in jax on amd gpus,'' \url{https://rocm.blogs.amd.com/artificial-intelligence/nanoGPT-JAX/README.html}, July 2024, accessed: 2024-09-07.

\bibitem{song2024mamba}
S.~Song, J.~Adeem, and M.~Arseny, ``Mamba on amd gpus with rocm,'' \url{https://rocm.blogs.amd.com/artificial-intelligence/mamba/README.html}, June 2024, accessed: 2024-09-07.

\bibitem{wang2024black}
X.~Wang, Y.~Li, Y.~Li, and G.~Kish, ``From black box to clarity: Ai-powered smart grid optimization with kolmogorov-arnold networks,'' \emph{arXiv preprint arXiv:2408.04063}, 2024.

\bibitem{Mughees_2024}
\BIBentryALTinterwordspacing
M.~Mughees, Y.~Li, and R.~Y. Li, ``From c.elegans to liquid neural networks: A robust wind power multi-time scale prediction framework,'' Aug. 2024. [Online]. Available: \url{http://dx.doi.org/10.36227/techrxiv.172469941.17523365/v1}
\BIBentrySTDinterwordspacing

\bibitem{li2022smart}
Y.~R. Li, F.~Nejabatkhah, and H.~Tian, \emph{Smart hybrid AC/DC microgrids: power management, energy management, and power quality control}.\hskip 1em plus 0.5em minus 0.4em\relax John Wiley \& Sons, 2022.

\bibitem{wang2013two}
P.~Wang, P.~McCluskey, and A.~Bar-Cohen, ``Two-phase liquid cooling for thermal management of igbt power electronic module,'' \emph{Journal of Electronic Packaging}, vol. 135, no.~2, p. 021001, 2013.

\bibitem{statespaces2024mamba}
S.~Spaces, ``Mamba: An efficient transformer model,'' \url{https://github.com/state-spaces/mamba}, 2024, accessed: 2024-09-07.

\bibitem{eleutherai2024gptneo}
EleutherAI, ``Gpt-neo: An implementation of gpt architecture by eleutherai,'' \url{https://www.eleuther.ai/artifacts/gpt-neo}, 2024, accessed: 2024-09-07.

\bibitem{nvidia2024v100}
NVIDIA, ``Nvidia volta v100 datasheet,'' \url{https://images.nvidia.com/content/technologies/volta/pdf/volta-v100-datasheet-update-us-1165301-r5.pdf}, 2024, accessed: 2024-09-07.

\bibitem{nvidia2024a40}
------, ``Nvidia a40 datasheet,'' \url{https://images.nvidia.com/content/Solutions/data-center/a40/nvidia-a40-datasheet.pdf}, 2024, accessed: 2024-09-07.

\bibitem{technicalcity2024comparison}
T.~City, ``Geforce rtx 4090 vs radeon rx 7900 xtx comparison,'' \url{https://technical.city/en/video/GeForce-RTX-4090-vs-Radeon-RX-7900-XTX}, 2024, accessed: 2024-09-07.

\end{thebibliography}
